\documentclass[pdflatex,sn-mathphys-num]{sn-jnl}
\usepackage{graphicx}%
\usepackage{multirow}%
\usepackage{amsmath,amssymb,amsfonts}%
\usepackage{amsthm}%
\usepackage{mathrsfs}%
\usepackage[title]{appendix}%
\usepackage{xcolor}%
\usepackage{textcomp}%
\usepackage{manyfoot}%
\usepackage{booktabs}%
\usepackage{algorithm}%
\usepackage{algorithmicx}%
\usepackage{algpseudocode}%
\usepackage{listings}%
\usepackage{hyperref}
\usepackage{bm}
\usepackage{changes}
\usepackage{float}
\usepackage{tabularx}
\newcommand{\bse}{\begin{subequations}}
	\newcommand{\ese}{\end{subequations}}

\renewcommand{\ss}[1]{_{\hbox{\tiny #1}}}

\def\bfi{\begin{figure}[H]}
	\def\efi{\end{figure}}

\newcommand{\be}{\begin{equation}}\newcommand{\ee}{\end{equation}}
\newcommand{\bea}{\begin{eqnarray}}\newcommand{\eea}{\end{eqnarray}}
\newcommand{\brr}{\begin{array}}\newcommand{\err}{\end{array}}
\newcommand{\bit}{\begin{itemize}}\newcommand{\eit}{\end{itemize}}
\newcommand{\ben}{\begin{enumerate}}\newcommand{\een}{\end{enumerate}}

\newcommand{\bbm}{\begin{bmatrix}}\newcommand{\ebm}{\end{bmatrix}}
\newcommand{\ba}{\begin{array}}
	\newcommand{\ea}{\end{array}}

\newcommand{\ble}{\begin{Lemma}} \newcommand{\ele}{\end{Lemma}}

\def\ha{\frac{1}{2}}

\def\al{\alpha}

\def\1{{_{1}}}\def\2{{_{2}}}

\def\noHe0{:\;\!\!\;\!\!:H_e(0):\;\!\!\;\!\!:}
\def\noHm0{:\;\!\!\;\!\!:H_\mu(0):\;\!\!\;\!\!:}
\def\al{\alpha}

\def\1{{_{1}}}\def\2{{_{2}}}

\definecolor{darkred}{rgb}{.8,0,0}

\definecolor{darkblue}{rgb}{0,0,.7}

\begin{document}
	
	\title{Coarsening kinetics in spin systems with long-range interactions: from voter to Ising}

	\author[1,2]{\fnm{Federico} \sur{Corberi}}\email{fcorberi@unisa.it}
	\author[3]{\fnm{Eugenio} \sur{Lippiello}}\email{eugenio.lippiello@unicampania.it}
	\author[4,5]{\fnm{Paolo} \sur{Politi}}\email{paolo.politi@cnr.it}
	\author[1,2]{\fnm{Luca} \sur{Smaldone}}\email{lsmaldone@unisa.it}
	
	\affil[1]{\orgdiv{Dipartimento di Fisica}, \orgname{ Universit\`a di Salerno}, \orgaddress{\street{Via Giovanni Paolo II, 132}, \city{Fisciano}, \postcode{84084}, \country{Italy}}}
	\affil[2]{\orgdiv{INFN Sezione di Napoli}, \orgname{ Gruppo collegato di Salerno}, \country{Italy}}
	\affil[3]{\orgdiv{Dipartimento di Matematica e Fisica}, \orgname{Universit\`a della Campania}, \orgaddress{\street{Viale Lioncoln, 5}, \city{Caserta}, \postcode{81100}, \country{Italy}}}
	\affil[4]{\orgdiv{Istituto dei Sistemi Complessi}, \orgname{Consiglio Nazionale delle Ricerche}, \orgaddress{\street{via Madonna del Piano, 10}, \city{Sesto Fiorentino}, \postcode{I-50019}, \country{Italy}}}
	\affil[5]{\orgdiv{INFN Sezione di Firenze}, \orgaddress{\street{via G. Sansone, 1}, \city{Sesto Fiorentino}, \postcode{I-50019}, \country{Italy}}}

	\abstract{
		In this paper, we start reviewing the main features of the one-dimensional Ising model with long-range interactions, where the spin-spin coupling decays as a power law, $J(r) \propto r^{-\alpha}$. We then discuss the key properties of the one-dimensional voter model, in which two agents (spins) at distance $r$ interact with a power-law probability with the same form of $J(r)$. The two models are compared, and the so-called $p$-voter model is presented, which provides a framework to interpolate between them. Specifically, the $p$-voter model reduces to the voter model for $p = 1$ and $p = 2$, while for $p \ge 3$ it falls into the universality class of the Ising model.
	}
	
	
	%%\pacs[JEL Classification]{D8, H51}
	
	%%\pacs[MSC Classification]{35A01, 65L10, 65L12, 65L20, 65L70}
	
	\maketitle

\section{Introduction} \label{SecIntro}

One of the most studied and paradigmatic classes of models in statistical mechanics 
is that of
spin models ~\cite{Huang1987,Pathria2011,Baxter1982,krapivsky2010kinetic,LiviPoliti2017}. One of their major achievements has been to provide deep insight into the fundamental mechanisms underlying phase transitions and kinetic behaviors. Despite their drastic simplification of real systems, these models are capable of capturing the essential ingredients responsible for the emergence of collective behavior. This many-body nature becomes even more pronounced when the interaction is not restricted to nearest neighbors (nn), as in the simplest formulations, but extends over the entire system via long-range couplings~\cite{BrayRut94,RutBray94,CMJ19,Corberi_2017,Corberi2019JSM,PhysRevE.103.012108,Corberi2021SCI,CORBERI2023113681,Corberi2023PRE,PhysRevE.102.020102,CorbCast24,corsmal2023ordering,corberi2024aging,corberi20243d}.

A particularly interesting topic in the nonequilibrium domain is the study of \emph{phase ordering}~\cite{Bray94,krapivsky2010kinetic,LiviPoliti2017}, which refers to the dynamical process by which order emerges in a system quenched from a high-temperature homogeneous phase to a low-temperature one with broken symmetry. This process occurs via the formation and coarsening of domains, whereby regions of distinct equilibrium phases grow over time. The characteristic length scale of these domains, \( L(t) \), increases as a function of the time elapsed after the quench.

In particular,
the nn Ising model (IM) has been extensively studied since the early days of statistical mechanics both for its equilibrium \cite{Lenz1920,Ising1925,PhysRev.65.117,RevModPhys.39.883} and non-equilibrium \cite{Glauber} properties. In the case of non-conserved dynamics, the system evolves through spin-flip processes
regulated by transition rates obeying detailed balance (e.g. Metropolis, Glauber, or heat bath rules), allowing it to reach the ground state. 
In one dimension ($D=1$), the dynamics can be effectively described in terms of random walks performed by domain walls (DWs). Consequently, it is not surprising that the average domain size grows according to the well-known law~\cite{Glauber} \( L(t) \simeq t^{1/2} \), a feature that remains unchanged also in 
higher dimensions. The situation is actually more complicated when long-range interactions are introduced. For example, when the coupling between different spin sites has a power-law form, 
$J(r) \propto r^{-\al}$, one finds that, varying $\al$, the system presents a different phase ordering dynamics~ \cite{CLP_review,Corberi2021SCI,Corberi_2019,Corberi_2017,BrayRut94,RutBray94}. 

Another spin model exhibiting phase ordering is the voter model (VM) ~\cite{Clifford1973, Holley1975, liggett2004interacting, Theodore1986, Scheucher1988, PhysRevA.45.1067, Frachebourg1996, Ben1996, PhysRevLett.94.178701, PhysRevE.77.041121, Castellano09,krapivsky2010kinetic,yjf2-4z1d}. This model, that was initially introduced to describe opinion dynamics and ecological systems, has the advantage of being exactly solvable. This is because, while in the IM spin updates are determined by the local field generated by the other spins, in the VM a randomly chosen agent copies the state of a nn chosen at random. Therefore, a spin flip attempt is determined by the state of a unique other agent.
As a consequence, with nn interactions the VM is equivalent to the IM in one dimension, but not for $D>1$. In general the system reaches the consensus state (where all spins are aligned) through a coarsening process, characterized by an average domain size growing as $L(t) \propto t^\ha$, for $D \leq 2$, while for $D>2$ the coarsening process reaches a metastable stationary state. 
Besides its many application in social sciences and other interdisciplinary fields~ \cite{Zillio2005,Antal2006,Ghaffari2012,CARIDI2013216,Gastner_2018, Castellano09,Redner19}, the VM represents also a 
sort of simplified setting where some phase ordering features can be better understood, due to its analytical tractability. 

A series of recent works~\cite{CorbCast24,corsmal2023ordering,corberi2024aging,corberi20243d} have generalized the analytical study of the VM~\cite{Frachebourg1996,Ben1996} to include long-range interactions. In this version of the VM, a randomly selected spin \( S_i = \pm 1 \), located at site \( i \) of a lattice, aligns with another spin at a distance \( r \), chosen with a probability \( P(r) \propto r^{-\alpha} \). With this rule, the VM can be regarded as a proxy for the IM with long-range power-law interactions $J(r)\propto r^{-\alpha}$. It is found that, varying $\al$, the system presents rather different properties, that will be summarized, limiting to the one-dimensional case, in this paper. %In particular, for $\al$ small %enough (i.e. for long-range %interactions), the system reaches a %stationary state (different from the %consensus state) even for $D \leq %2$, while for $\al \leq D$ there is %not a macroscopic coarsening process %and the system immediately %approaches such stationary states. 
An hybrid version of the VM, where the interaction could be both of long-range or nn nature, with a given probability, has been also studied and exactly solved in Ref. \cite{GLR25}.

In Ref.~\cite{Corberi_2024}, the \( p \)-voter model (pVM) was introduced. In this model, a spin aligns with the majority of \( p \) other spins, randomly selected at distances \( r \) from \( i \), with the same power-law probability distribution of the VM with long-range interactions mentioned above. 
%A similar model, though of mean-field character (\( \alpha = 0 \)), was previously introduced and studied in~\cite{PhysRevE.92.052105,Chmiel2018}. 
The pVM interpolates between the long-range voter model (recovered for \( p = 1 \)) and the long-range Ising model with coupling \( J(r) = P(r) \) %and $T=0$ 
in the limit \( p \to \infty \). Moreover, for \( p \geq 3 \) it enters the universality class of the Ising model. 
Thus, for finite but sufficiently large \( p \), the pVM can be seen as an effective and computationally tractable approximation of the IM, replacing its all-to-all coupling with a \( p \)-body interaction.

In this paper, we review the main features of the one-dimensional IM with long-range interactions, focusing on phase-ordering kinetics.
This will be done in Section~\ref{IMsec}. 
The reason to consider $D=1$ is because this is the case where the kinetics is better understood, because some approximate analytical techniques are available and numerical simulations are more easily affordable \cite{CLP_review}. Then, in Section~\ref{SecModels}, we examine the one-dimensional VM \cite{CorbCast24} with analogous power-law interaction probabilities. The hybrid VM \cite{GLR25} is also very briefly discussed. After a comparison of the IM and the VM (Section~\ref{sec_compa}), we review the pVM \cite{Corberi_2024}, which serves as a unifying framework bridging these two models, in Section~\ref{pVMsec}.
This work aims to provide a concise and schematic overview of the ordering kinetics in one-dimensional systems with long-range interactions, drawing upon a range of recent results and developments~\cite{Bray94,CLP_review,CMJ19,Corberi2021SCI,Corberi_2019,Corberi_2017,PhysRevE.103.012108,BrayRut94,RutBray94}.

%The paper is organized as follows: %in Section \ref{IMsec}, main facts %on the ordering kinetics of the IM %model are reviewed. In Section %\ref{SecModels} the main results on %the VM are also reported, while in %Section \ref{pVMsec} the pVM %coarsening behavior is briefly %reviewed. 

\section{IM with long-range interactions} \label{IMsec}

\subsection{The model}

We consider a general one-dimensional IM described by the Hamiltonian 
\be
{\cal H} = -\sum_{i} \sum_{r>0} J(r) S_i S_{i+r} ,
\label{HIsing}
\ee 
where $S_i=\pm 1$, $J(r)$ is a positive, non increasing function, and the lattice is formed by $N$ sites,
with periodic boundary conditions.
The usual Ising Hamiltonian couples only nn spins,
$
J\ss{nn}(r)=\delta_{r,1}
$. In this work we will focus on the case of power-law long-range interactions
\be
J(r) = r^{-\al} ,
\ee
for $r\ge 1$, which reduces to $J\ss{nn}(r)$ when $\al \to \infty$.
Mean field theory corresponds to $\alpha=0$.
The equilibrium properties of the one-dimensional long-range IM are well known and it is useful to summarize them below.

The model has no phase transition if $\alpha>2$ because the cost $E_{dw}$ of a single DW (which, locating it at the origin for simplicity, is $E_{dw}=2\sum_{i=-\infty}^0 \sum_{j=1}^{\infty} J(j-i)$), is finite , as for the nn model.
If $\alpha\le 1$, the cost $E_{sf}$ of a single spin flip diverges
($E_{sf} = 4\sum_{i=1}^\infty J(i)$):
we are in the \emph{strong long-range} (SLR) regime, characterized by additivity breaking.
If $\alpha\le 2$ there is a phase transition~\cite{Dyson1969} 
with mean-field critical exponents for $\alpha\le 3/2$~\cite{Mukamel2009}.
Finally, if $\alpha=2$ there is a mixed-order phase transition~\cite{Bar2014}, because the critical point is
characterized by a discontinuity of the magnetization~\cite{Aizenman1988} 
as well as by a diverging correlation length~\cite{Cardy1981}.

In this review we focus on the so called non-conserved order parameter dynamics, which is based on single spin-flip processes, $S_k \to -S_k$, also called Glauber dynamics.
This rule will be implemented using
the Glauber transition rate, according to which the 
probability per unit time to 
move from an initial configuration 
of energy $E_I$ to another of
energy $E_F$
%transition probability between
%a general initial configuration of %energy $E_I$ and a final %configuration of energy $E_F$
is
\be
W\ss{IF} = \frac{1}{1+e^{\beta(E_{F} -E_{I})}} 
\label{glauber}
\ee
where $\beta$ is the inverse temperature (in units of the Boltzmann constant $k_B$).
For the Hamiltonian (\ref{HIsing}), if we define the Weiss field acting on site $i$,
$H_i = \sum_{j\ne i} J(|i-j|) S_j$, Eq.~(\ref{glauber}) reduces to
\be
W\ss{IF}(S_i)=\frac{1}{2}[1-S_i \tanh (\beta H_i)],
\label{transT}
\ee
where the dependence of $W\ss{IF}$ on $S_i$ indicates that the $i-$th spin is attempting to flip.
For a vanishing temperature, $\beta\to \infty$, the transition rate is further simplified,
\be
W\ss{IF}(S_i)=\frac{1}{2}[1-S_i \mbox{sign}(H_i)],
\label{transIsing}
\ee
with $\mbox{sign}(0) = 0$.

%%%%%%%%%%%%%%%%%%%%%%%%%%%%%%%%%%%%%%%%%%%%%%%%%%%%%%%%%%%%%%%%%%%%%%

\subsection{Analytical predictions and numerical results} \label{simplified}

The kinetics of the IM with algebraic interactions cannot be exactly solved. However it has been  
studied numerically and with approximate analytic approaches in~\cite{CLP_review,Corberi2021SCI,Corberi_2019,Corberi_2017,BrayRut94,RutBray94} as we will shortly review, mainly following Ref.~\cite{CLP_review}.

After a quench from the fully disordered phase, i.e. with $\langle S_i\rangle=0$ and $\langle S_iS_j\rangle=\delta_{ij}$, which corresponds to an initial temperature $T_i=\infty$, to zero or low temperature,
the system relaxes to equilibrium, which is characterized by a local order
on a length scale equal to the equilibrium correlation length $\xi(T)$. 
Upon lowering $\alpha$, the correlation length increases as well
and we even have long range order at finite $T$, if
$\al \le 2$. In any case, even if $T\ss{c}=0$ we can consider low enough $T$ such that $\xi(T)$
is arbitrarily large.

Relaxation to equilibrium occurs through a coarsening kinetics where domains disappear leading to 
the increase over time of their average size, $L(t)$, one possible
definition of which is
\be
L(t)=\frac{\sum _{r=0}^{N/2}r\,C(r,t)}{\sum _{r=0}^{N/2}C(r,t)} \, , 
\label{eqL}
\ee
where $C(r,t) = \langle S_i(t) S_{i+r}(t)\rangle - \langle S_i(t)\rangle\langle S_{i+r}(t)\rangle$ 
is the two-point correlation function which is independent on $i$ due to space translation invariance.
Another definition (more natural and easier to implement in $d=1$) is the average distance between DWs.
Therefore, if $\rho(t)$ is their density, the average domain length can be estimated as $\rho^{-1}(t)$. In the present model the two definitions give similar results, $L(t) \propto \rho^{-1}(t)$.
In the next Section on the VM we will see that this coincidence does not generally occur.

$L(t)$ can be easily determined numerically,
but its analytical derivation generally requires strong simplifications.
Actually, if $\alpha >1$ the energy of a single domain
is finite and the determination of its closure time is all we need:
this is because
the evolution is a self-similar phenomenon characterized by a single length scale (i.e. $L(t)$).
According to the scaling hypothesis, this property means, e.g., that
$C(r,t)$
is a function of a single variable,
\be
C(r,t) = f(x), \qquad \mbox{with } x \equiv \frac{r}{L(t)}.
\label{scalform}
\ee

Invoking scaling, it can be argued~\cite{CLP_review} that the functional dependence $L(t)$ can be found 
by determining the typical temporal scale $t$ necessary to close a domain of initial size $L$
and inverting the resulting function $t(L)$.

\bfi
\begin{center}
	\vspace{-2cm}
	\rotatebox{270}{\resizebox{.6\textwidth}{!}{\includegraphics{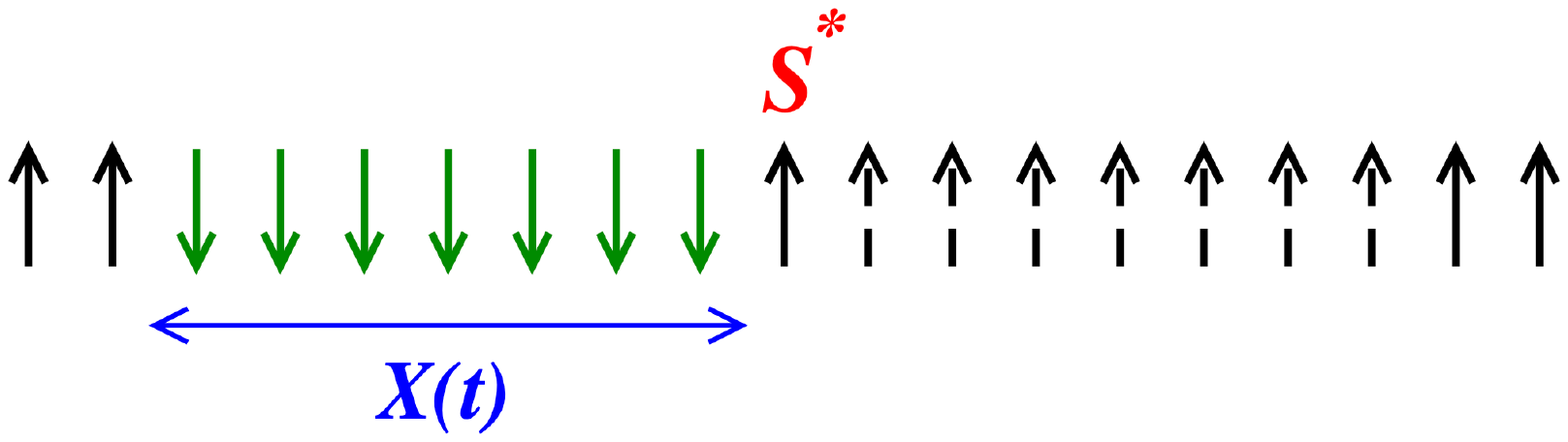}}}
	\vspace{-2cm}
\end{center}
\caption{
	Simple one dimensional configuration, with periodic boundary conditions,
	with a single domain of down spins and length $X(t)$ (with $X(0)=L$).
	See the main text for further details.}
\label{fig.models}
\efi

Let us consider a domain of the negative phase and initial size $X(0)=L$,
see Fig.~\ref{fig.models}, surrounded by a sea of positive spins. In the nn case, 
each interface performs a symmetric random walk 
because the energy of a given configuration depends on the number of DWs only,
not on their positions.
Introducing a long-range interaction, DWs interact with each other, implying  that an interface
now performs an anisotropic random walk. For the simple, one-domain configuration
of Fig.~\ref{fig.models} there is a drift favoring the closure
of the domain $X(t)$, whose time we are now going to evaluate.

The key point is that the interaction of spin $S^*$
with down spins is compensated by its interaction with dashed up spins.
For this reason it is useful to define the integrated quantity
\be
I(X) \equiv \sum_{r=X}^\infty J(r) .
\label{eq.I}
\ee
The process $X\to X+1$ requires the energy $(\Delta E)_+ = 4I(X)$,
while the process $X\to X-1$ releases the same energy.\footnote{More precisely, it is the process
	$X+1 \to X$ to release the same energy, but for large $X$ we can neglect this difference.}

Therefore, using Eq.~(\ref{glauber}), the probabilities of such processes are
\be
p_\pm = \frac{1}{1 + e^{\pm 4\beta I(X)}},
\ee
and the drift is proportional to the asymmetry,
\be
\delta(X) = p_+ - p_- = - \tanh (2\beta I(X)),
\label{exprdelta}
\ee
the negative sign indicating that the drift favors the closure of the domain.

Upon mapping the discrete, asymmetric random walk onto a convection-diffusion equation 
with diffusion coefficient $D$, and 
constant drift $v$ (because of scaling, $v$ is assumed to depend on $X(0)$, not on $X(t)$),
the average closure time of the particle, given in~\cite{bookRedner}, is
\be
t(L) = \frac{L}{v} \tanh\left( \frac{vL}{D} \right). 
\label{tL}
\ee
This expression gives the correct limits $t(L)=L^2/D$ for vanishing drift and $t(L)=L/v$ for strong drift
($v\gg D/L$). According to the spirit of the above calculation and in view of Eq.~(\ref{exprdelta}),
the appropriate expression to be used for the drift is 
\be
v(L) = v_0\tanh (2\beta I(L)), 
\label{v}
\ee
where $v_0$ is a constant and $I(L)$ can be evaluated in the continuum approximation, giving
\bea
I(L) \simeq \frac{1}{\al-1} \frac{1}{L^{\al-1}} \; . \label{I}
\eea
In conclusion, Eqs.~(\ref{tL}-\ref{I}) give the closing time $t$ of a domain as a function of its initial size $L$.
Inverting this relation the growth law $L(t)$ can be determined.

%By means of the above simplified model and thanks to numerical simulations, it can be proved the existence of 
%three regimes:
%\begin{itemize}
%\item 
%	For $\al> \al_{SR}$ (with $\al_{SR}=2)$, we are in the \emph{short-range} (SR) regime.
%\item
%	For $\al_{LR} < \al \le \al_{SR}$ (with $\al_{LR}=1)$, we are in the \emph{weak long-range} (WLR) regime.
%\item
%	For $\al \le \al_{LR}$, we are in the SLR regime.
%\end{itemize}
%
%If we are only interested in the asymptotic stage, $t,N\to\infty$, dynamics would be simple
%because the coarsening exponent, defined by $L(t)\propto t^n$, is $n=1/2$ in the SR and $n=1/\alpha$ in the WLR.
%In the SLR, mean-field dynamics prevents the formation of a multi-domain state and coarsening does not occur
%even in the thermodynamic limit.
%In fact, the study of the dynamics at large but finite times and system sizes gives richer
%and more interesting scenarios, which we are now going to briefly illustrate,
%referring the reader to the original articles for further details.

By means of the above simplified model the kinetic behavior of the long-range IM has been studied in~\cite{CLP_review}.
The outcome has been shown to compare well with  numerical simulations. Below,
we provide a summary of what is found.

A first clear distinction should be made between $\alpha$ smaller or larger than one.
In the latter case, the system is additive and we can assume a diverging system size 
without difficulty. Relaxation dynamics proceeds via standard coarsening whose law $L(t)$
can be found using Eqs.~(\ref{tL})-(\ref{I}).
The power-law coupling induces an attractive drift between neighboring DWs:
At relatively short times/domain sizes, the drift is constant and the domain closes ballistically, i.e., for the growth exponent $a$ defined by $L(t)\propto t^a$, one finds
$a = 1$.
At larger times, the drift decreases with the distance, giving an exponent
$a(\alpha)=1/\alpha < 1$. This behavior is the asymptotic one for $\alpha \le 2$ but transient for $\alpha >2$. In the latter case, at  even larger times the drift becomes vanishingly small,
reinstating a purely unbalanced diffusion of DWs and giving $a=1/2$,
as for the nn case.
The crossovers between these regimes depend on temperature, of course.
In particular, for vanishing $T$, since there is no thermal energetic scale to compare with, the strength of the drift is irrelevant and the coarsening is constantly ballistic. Numerical results, which show the above mentioned behavior, are reported in Figs. \ref{fig_Lt_pw_NCOP} and \ref{fig_Lt_sigmas_NCOP}.

For $\alpha <1$, the system is not additive and the system size enters directly to determine
the relaxation dynamics: in the
thermodynamic limit mean-field dynamics prevents the formation of a multi-domain state and coarsening does not occur.
Instead, a system of finite size $N$ has a probability ${\cal P}_\alpha(N)$ to relax via coarsening
and a probability $(1-{\cal P}_\alpha(N))$ to display mean-field behavior.
The peculiar nature of the regime $\alpha<1$ emerges when we realize that the two limits
$\alpha\to 1$ and $N\to\infty$ are not interchangeable:
$\lim_{\alpha\to 1}\lim_{N\to\infty}{\cal P}_{\alpha}(N) = 0$ while $\lim_{N\to \infty}\lim_{\alpha\to 1}{\cal P}_{\alpha}(N) = 1$ \cite{Corberi2021SCI}.

In conclusion, focusing in particular on the asymptotic stages, one can identify three $\alpha$ regimes:
\begin{itemize}
	\item 
	For $\al> \al_{SR}$ (with $\al_{SR}=2)$, we are in the \emph{short-range} (SR) regime. The exponent $a$ takes the same value as for nn.
	\item
	For $\al_{LR} < \al \le \al_{SR}$ (with $\al_{LR}=1)$, we are in the \emph{weak long-range} (WLR) regime, with an $\alpha$ dependent exponent $a$.
	\item
	For $\al \le \al_{LR}$, we are in the SLR regime, where non additivity and mean field behaviors show up.
\end{itemize}

\bfi
\centering
\rotatebox{0}{\resizebox{.85\textwidth}{!}{\includegraphics{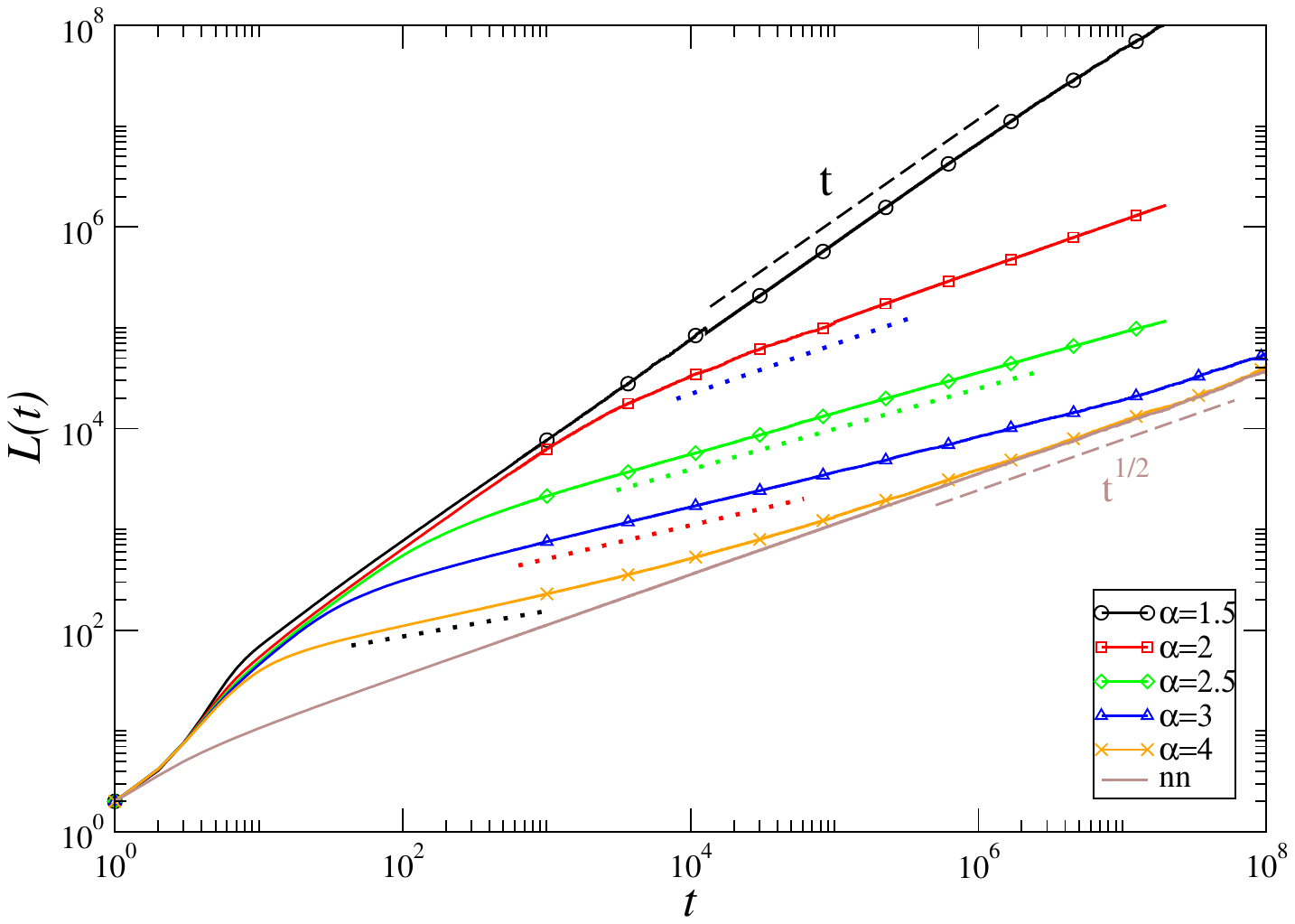}}}
\caption{
	$L(t)$ for a system quenched from $T_i=\infty$ to $T=10^{-3}$ on a double-logarithmic scale.
	Different symbols and colors
	correspond to different values of
	$\al$ and to the nn case (see legend).  The dashed orange line
	is the $t^{1/2}$ law and the dashed green one is the ballistic behavior.  
	The color dotted lines (below the data curves) are the power-laws
	$t^{1/\al}$ for each $\al$ value \cite{CLP_review}.
}
\label{fig_Lt_pw_NCOP}
\efi

\bfi
\centering
\rotatebox{0}{\resizebox{.75\textwidth}{!}{\includegraphics{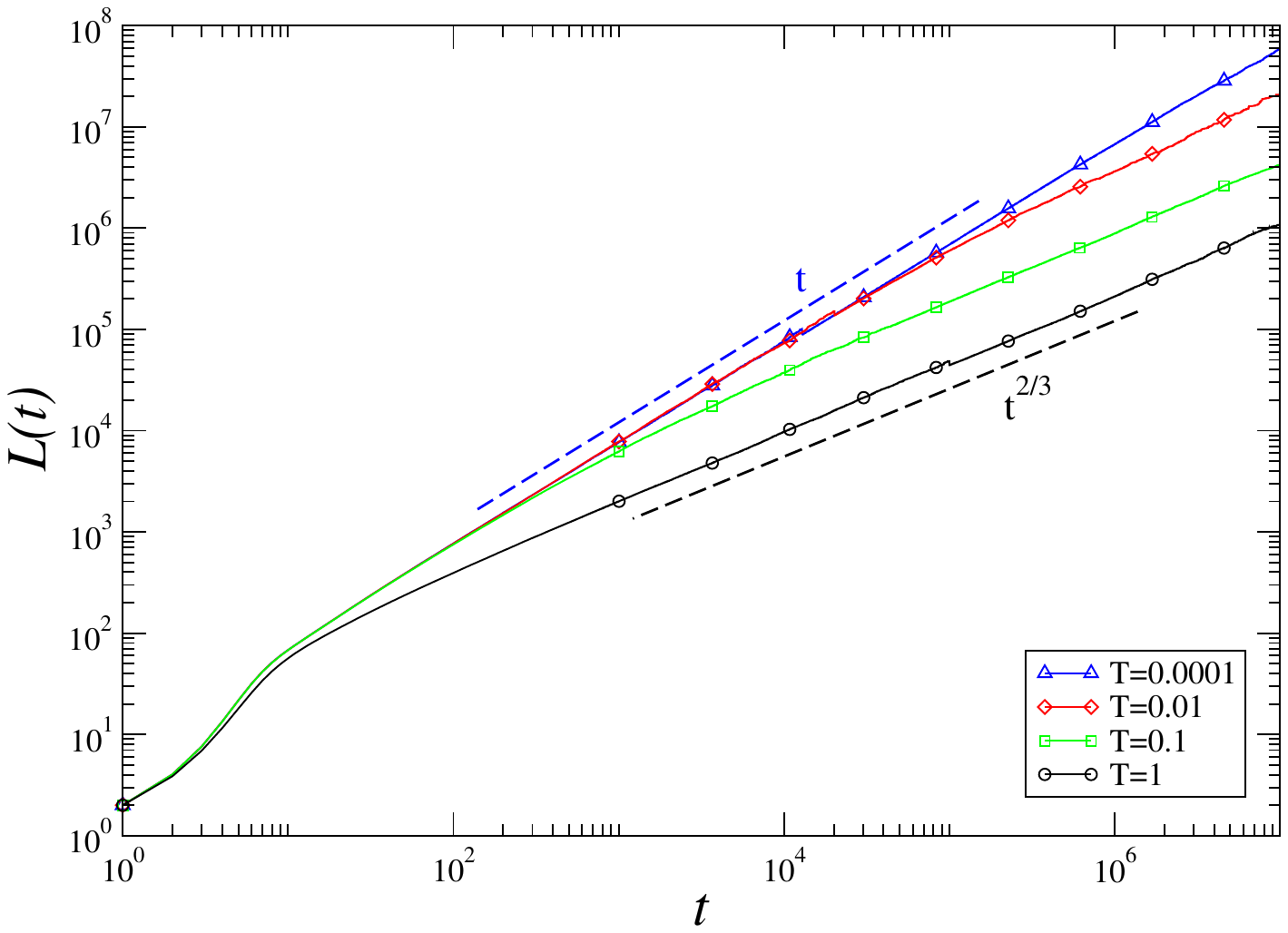}}}
\caption{
	$L(t)$ for a system quenched from $T_i=\infty$ to different final $T$ for $\al=1.5$.  The green dashed line 
	indicate the linear, ballistic regime $L(t)\sim t$.
	The magenta dashed line is the growth
	$L\sim t^{1/\al}=t^{2/3}$ in the slow regime.
	The system size is $N=8\times 10^8$ \cite{CLP_review}.}
\label{fig_Lt_sigmas_NCOP}
\efi

%%%%%%%%%%%%%%%%%%%%%%%%%%%%%%%%%%%%%%%%%%%%%%%%%%%%%%%%%%%%%%%%%%%%%%%%%%%%%%%%%%%%%%%%

\section{Voter model with long-range interactions}  \label{SecModels}

Analogously to the IM, the one-dimensional VM is defined in terms of spin variables $S_i=\pm 1$
(also called agents), where $i=1,\dots,N$
runs over the sites of a linear lattice with periodic boundary conditions. In the original nn version of the model an elementary move consists in aligning a randomly chosen spin to one of its nn, 
this one also chosen randomly.
In the long-range version, in an elementary move, a randomly chosen variable $S_i$ takes the state of another, $S_j$, 
chosen at distance $r$ with probability
$P(r)=\frac{1}{2Z}\,r^{-\alpha}$, where $Z=\sum_{r=1}^{N/2}r^{-\alpha}$. 
In one dimension, the case we will focus on in this paper, the probability to flip $S_i$ in a single update can thus be written~\cite{CorbCast24} as
\be
W_{VM}(S_i) \ =\ \frac{1}{2N}\sum _{r=1}^{N/2}P(r)\sum _{k=[[i\pm r]]}(1-S_iS_k).
\label{transrates}
\ee
The symbol $[[\dots]]$
means that the distance between spins must be correctly 
evaluated, because of periodic boundary conditions:
$[[i+r]]=i+r-N$ if $i+r>N$.
For an $n$-spin correlator it is easy~\cite{Glauber} to obtain
$\frac{d}{dt}\langle S_{i_1}S_{i_2}\cdots S_{i_n}\rangle=-2\langle S_{i_1}S_{i_2}\cdots S_{i_n}\cdot \sum _{m=1}^n w_\alpha(S_{i_m})\rangle$
which, specified for $n=2$ gives~\cite{CorbCast24}
\be
\frac{d}{dt}C(r,t)=-2C(r,t)+2\sum_{\ell=1}^{N/2}P(\ell)\left [C([[ r-\ell]],t)+C([[r+\ell]],t)\right ],
\label{eqc2}
\ee
where time is measured in Monte Carlo steps, and we assumed space translation invariance.
From $C(r,t)$ we can thus compute the correlation length through Eq.~\eqref{eqL}.

In Fourier space we obtain
\begin{align}\label{eq:differential_equation_fourier}
	\frac{d}{dt} C(q,t) &= \mathcal{A}(q) \, C(q,t) + v(t),
\end{align}
where
\begin{equation}
	\mathcal{A}(q) = 4 \, \sum_{\ell=1}^{N/2} P(\ell) \cos(\ell q) \, - \, 2 \, ,
\end{equation}
and $v(t)$ is a time-dependent Lagrange multiplier introduced to enforce the constraint $C(0,t)=1$ at all times.  

The formal solution of~\eqref{eq:differential_equation_fourier} reads
\begin{align}\label{eq:formal_solution_fourier}
	C(q,t) = e^{\mathcal{A}(q)t} 
	+ \int_0^t dt' \, v(t') \, e^{\mathcal{A}(q)(t-t')},
\end{align}
where we used the initial condition that all spins are uncorrelated,  
\[
C(\ell,0) = \delta_{\ell 0}, \qquad \Rightarrow \qquad C(q,0) = 1.
\]

Assuming scaling for the correlation function $C(r,t)$, see Eq.~\eqref{scalform},
its Fourier transform takes the form
\begin{align}\label{eq:scaling_form}
	C(q,t) = L(t) \, f\!\left(q L(t)\right) \, .
\end{align}
From the definition \eqref{eqL} and the Riemann--Lebesgue lemma, it follows that at large distances (and then large time), the main contribution to $L(t)$ is given by the small $q$ behavior of $\mathcal{A}(q)$.
The latter can be obtained analytically~\cite{cannas}:
\begin{equation}\label{eq:Aq_smallq}
	\mathcal{A}(q) \approx -K_1 q^2 - K_\sigma q^{\alpha-1},
\end{equation}
where $K_1$, $K_\sigma$ are constants and $\al>1$. Then, when $\mathcal{A}(q) \propto q^{1/a}$, we have $L(t) \propto t^a$.

From Eq.~\eqref{eq:Aq_smallq} it is clear that the large-time behaviors of $L(t)$ changes when $\al$ is varied.
As in the IM, three distinct dynamical regimes emerge depending on the value of $\alpha$. 
However, the lower bound of the short-range region is now $\al_{SR}=3$. Moreover, for $\alpha \le 2$ a stationary state without consensus (namely full ordering) is entered. These features will be further discussed in the following paragraphs. Similar results, but with shifted values of $\al_{SR}$ and $\al_{LR}$ are obtained in higher dimension~\cite{corsmal2023ordering,corberi20243d}. 

Let us notice that the notion of $\al_{LR}$ we have given for the IM is related to the fact that, below that value, the interaction is no-longer summable. 
However, in the VM, which is not described by an Hamiltonian, such notion is quite arbitrary. This fact led to different conventions in literature \cite{Corberi_2024,corberi20243d}. Here we define $\al_{LR}=1$ as in the IM.

\subsection{Case \texorpdfstring{\boldmath{$\alpha > 3$}}{alpha > 3}}
\label{agt3}

\begin{figure}[ht]
	\vspace{1.0cm}
	\centering
	\rotatebox{0}{\resizebox{0.75\textwidth}{!}{\includegraphics{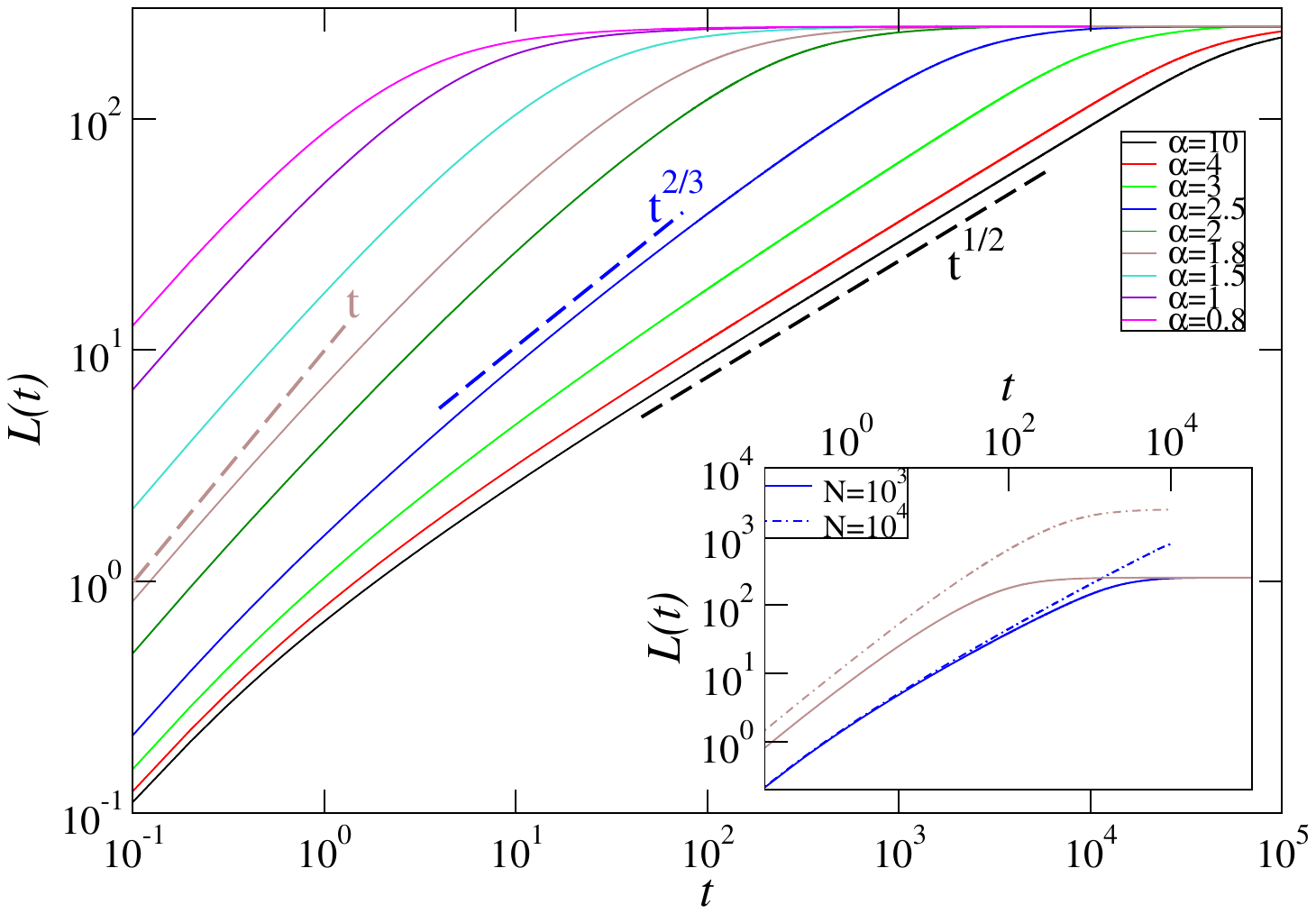}}} 
	\caption{$L(t)$ is plotted against time $t$, for different values of $\alpha$ (see legend). Data are obtained by solving numerically Eq.~(\ref{eqc2}) with $N=10^3$. 
		Dashed lines (with corresponding color as numerical data) are the analytical results of Eqs.~(\ref{diffgrowth},\ref{alphareg},\ref{balreg}). In the 
		inset a comparison is shown, for $\alpha=2.5$ and $\alpha=1.8$, with a larger system with $N=10^4$ (dot-dashed lines). Note that all curves saturate to $N/4$, which corresponds,  according to Eq.~\eqref{eqL}, to the fully ordered state \cite{CorbCast24}.}
	\label{figL}
\end{figure}

It is clear from Eq.~\eqref{eq:Aq_smallq} that for $\alpha>3$ the quadratic term dominates, yielding the nn behavior
\be
L(t) \ = \ D \,t^{1/2},
\label{diffgrowth}
\ee
where $D$ is a constant. For the density of DWs, in such case, one finds $\rho^{-1}(t) \propto L(t)$. This behavior is very clearly observed in Fig.~\ref{figL} where this quantity has been plotted after solving numerically Eq.~(\ref{eqc2}). After an initial transient, the curves for $\alpha >3$ show $L(t)$ growing as in Eq.~(\ref{diffgrowth}) before saturating when $L(t)$ approaches $N$,
when the finite size is felt and the system reaches consensus.

The correlation function, for $t \gg 1$ has the scaling form \eqref{scalform} and using
Eq.~(\ref{eqc2}) it is found to be 
\be
C(r,t) \ = \ \mbox{erfc} \left (\frac{x}{x_0} \right )  \, , 
\label{psiafter3}
\ee
where $x_0=\frac{2\langle \ell^2 \rangle^{1/2}}{D}$, 
$
\langle \ell^2 \rangle = \sum_{\ell=1}^{N/2}\ell ^2 P(\ell)
$
is the second moment of the distribution of interaction distances, and $\mbox{erfc}(x)=1-\mbox{erf}(x)$ is the complementary error function.

\begin{figure}[ht]
	\vspace{1.0cm}
	\centering
	\rotatebox{0}{\resizebox{0.75\textwidth}{!}{\includegraphics{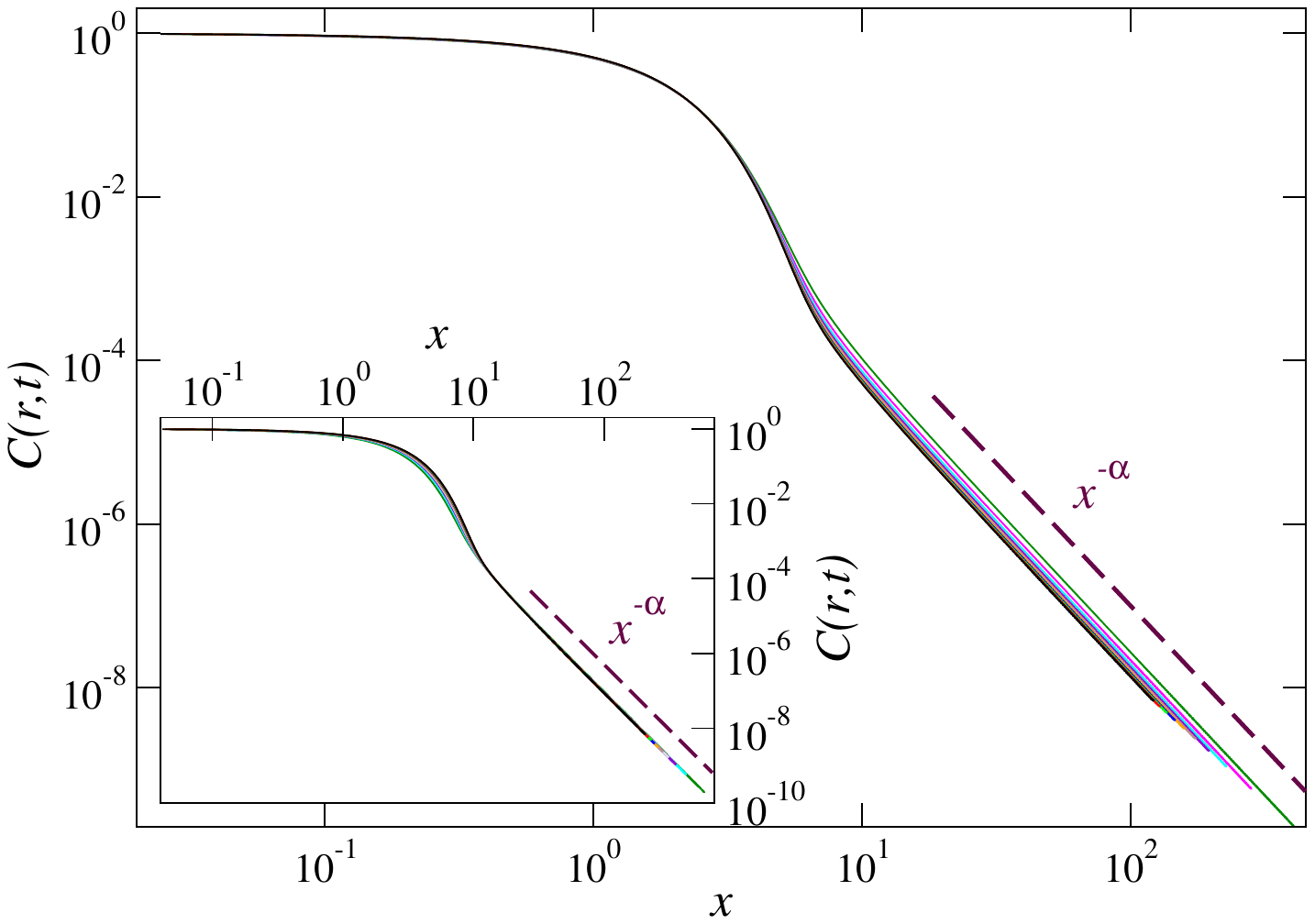}}} 
	\caption{Collapse plot, for $\alpha=3.5$, of the correlation function $C(r,t)$ as a function of $x=r/L(t)$ (main plot), or $x=r/\mathcal{L}(t)$ (inset). $L(t)$ is computed using Eq.~\eqref{eqL} and $\mathcal{L}(t)$ as defined in the text. System size is $N=10^4$. The curves are for values of $t$ ranging from $10^2$ to $10^3$ in steps of $10^2$ \cite{CorbCast24}.}
	\label{fig_C1}
\end{figure}

Strictly speaking, the above solution is only valid for $x$ smaller than a certain time-dependent value $x^*(t)$. $x^*(t)$ is an ever growing quantity as time goes by (it can be estimated that $x^*$ diverges logarithmically with time).  For finite time and $x>x^*(t)$, we have a different analytical behavior, $f(x) \ \propto x^{-\al}$ (this can be clearly observed in Fig.~\ref{fig_C1}), and a different coherence length 
$\mathcal{L}(t)\propto x_0^{1/\alpha}t^{3/(2\alpha)}$.
Due to the time dependence of $x^*(t)$ discussed above, it is clear that Eq.~(\ref{psiafter3}) is the correct behavior in the limit $t\to \infty$.

%%%%%%%%%%%%%%%%%%%%%%%%%%%%%%%%%%%%%%%%%%%%%%%%%%%%%%%%%%%%%%%%%%%%%%%%%%%%%%%%%%%%%%%%
\subsection{Case \texorpdfstring{\boldmath{$2 < \alpha \le 3$}}{2 < alpha <= 3}}
\label{ain23}

In this range of $\al$-values, the correlation function $C(r,t)$ (for $r\gg 1$) still obeys a scaling form as in Eq.\eqref{scalform}:
\be
C(r,t) \simeq \frac{1}{(ax)^\alpha},
\label{unomenxbis}
\ee
but with a scaling length
\be
L(t) \sim t^{1/(\alpha-1)} \, .
\label{alphareg}
\ee
In fact, the fractional term in Eq.~\eqref{eq:Aq_smallq} prevails in the present case.

\begin{figure}[ht]
	\vspace{1.0cm}
	\centering
	\rotatebox{0}{\resizebox{0.75\textwidth}{!}{\includegraphics{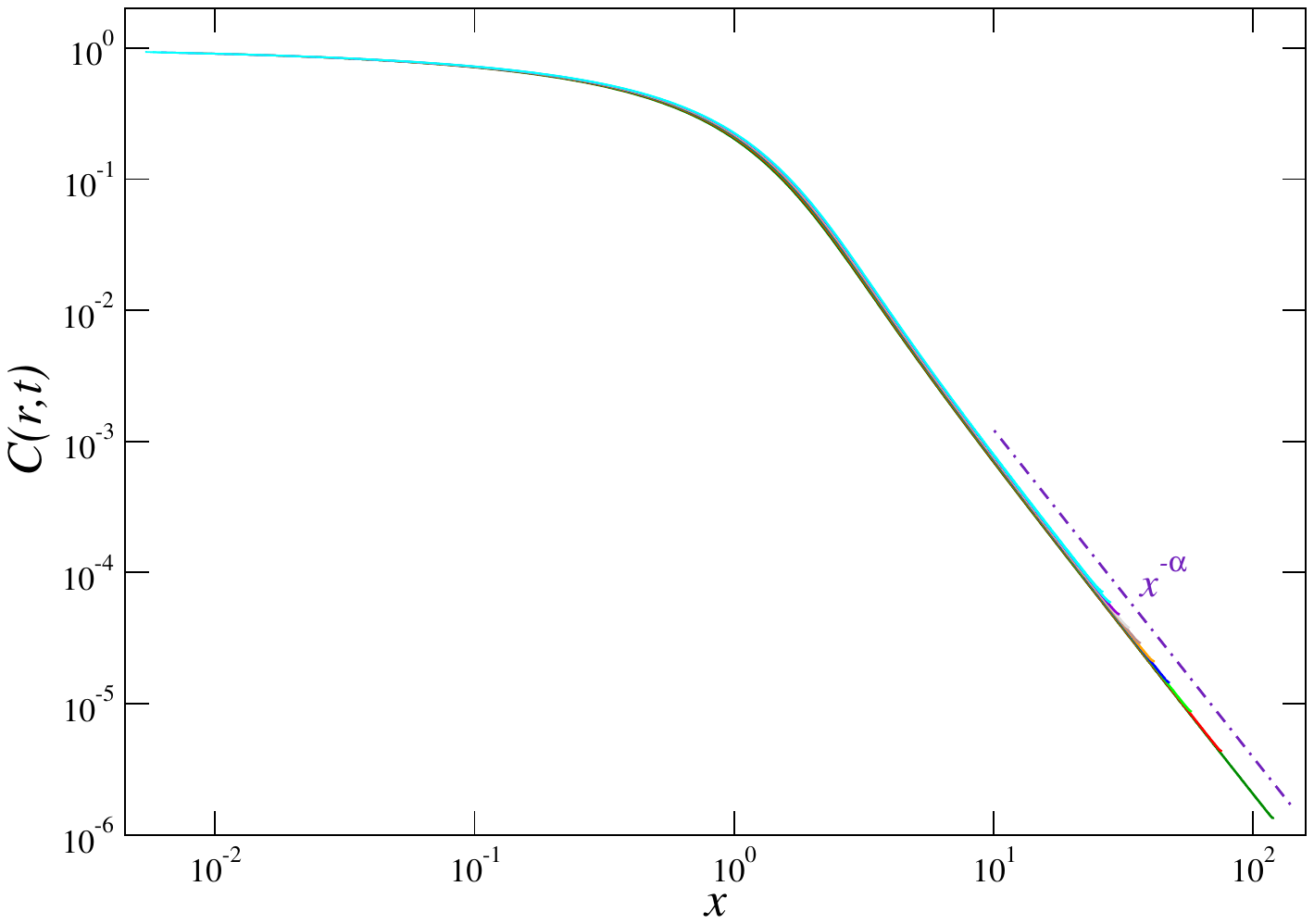}}}
	\caption{Collapse plot, for $\alpha=2.5$, of the correlation function $C(r,t)$ as a function of $x=r/L(t)$ where $L(t)$ is computed using Eq.~\eqref{eqL}. System size is $N=10^4$. The curves are for values of $t$ ranging (in steps of $10^2$ time units) from $10^2$ to $10^3$. The green dashed line is the behavior~(\ref{unomenxbis}) \cite{CorbCast24}. 
	}
	\label{fig_C2}
\end{figure}

The behavior~(\ref{alphareg}) is well reproduced by the numerical solution, as shown in Fig.~\ref{figL} for the case with $\alpha=2.5$. In addition, Fig.~\ref{fig_C2} confirms that scaling holds and that the
scaling function decreases as in Eq.~\eqref{unomenxbis} for large $x$.

Instead, the density of interfaces is 
\be
\rho^{-1}(t) \propto t^{(\alpha-2)/(\alpha-1)}.
\label{expdecayrhot}
\ee
Therefore, in this case $L(t)$ and $\rho^{-1}(t)$ are regulated by different exponents, signaling the presence of two distinct growing lengths. Physically what happens is that in a coherent region of size $L(t)$, inside which most spins are aligned in a given direction, a certain number of variables is reversed because, due to the long-range interaction, they can easily align with faraway agents belonging to domains of opposite sign. 
The typical distance between two of such reversed spins is $\rho^{-1}(t)\ll L(t)$.
Note also that the growth exponent in Eq.~(\ref{expdecayrhot}) matches the short-range value $1/2$ as $\alpha \to 3$,
making $\rho^{-1}(t) \sim L(t)$ in such limit.
Conversely, it vanishes as $\alpha \to 2$, signaling
that, for $\alpha \le 2$ the system reaches
a partially-ordered non-equilibrium stationary state, containing a finite amount of interfaces. This will be discussed in the next Section.

\subsection{Case \texorpdfstring{\boldmath{$0 \le \alpha \le 2$}}{0 <= alpha <= 2}}
\label{aless2}

By definition, a consensus state is one where all the agents share the same opinion. 
For $\alpha \le 2$ in the thermodynamic limit $N\to \infty$ such state is not reached, while it is finally achieved for any finite $N$ if we take the $t \to \infty$ limit first: the two limits do not commute. 
Therefore we here discuss the whole region $\al \leq 2$ in the same Section, although a further distinction occurs for $\al>1$ and $\al\le 1$, due to the passage from the WLR to the SLR regime.

When the thermodynamic limit is taken first, for $1< \alpha \le 2$ (WLR),
a non-trivial stationary state is attained, with 
\be
C_{stat}(r)=\kappa \,r^{-(2-\alpha)},
\label{Cstat}
\ee
for $r\gg 1$, where
$\kappa$ is a constant. Computing the correlation length, inserting this form into Eq.~(\ref{eqL}), one finds
\be
L_{stat}=\frac{\alpha-1}{2\alpha} \,N.
\label{Lstat}
\ee
This quantity diverges in the 
thermodynamic limit showing 
the absence of a typical length in such a state which is therefore, in some sense, critical. 

\begin{figure}[ht]
	\vspace{1.0cm}
	\centering
	\rotatebox{0}{\resizebox{0.49\textwidth}{!}{\includegraphics{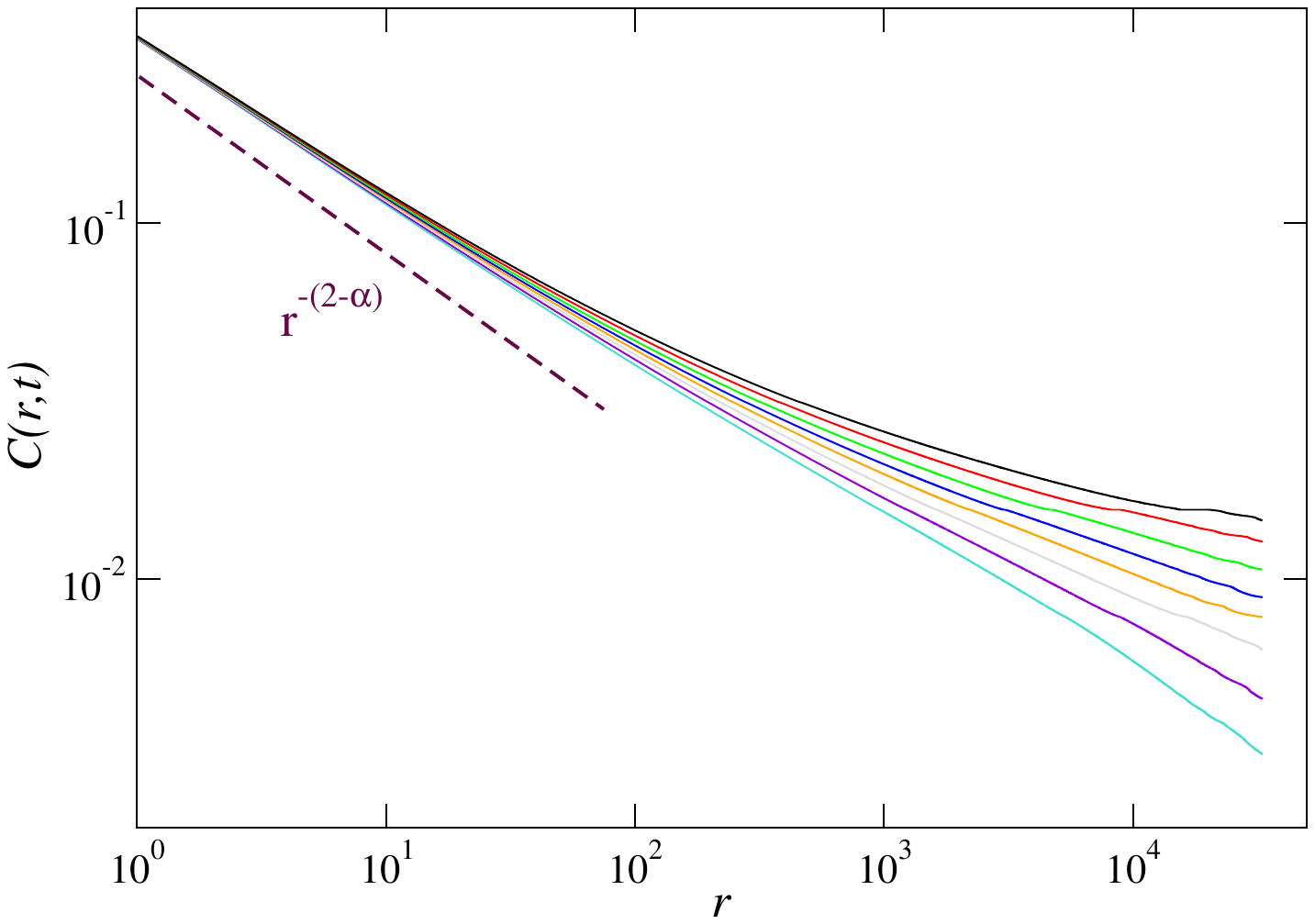}}}
	{\resizebox{0.49\textwidth}{!}
		{\includegraphics{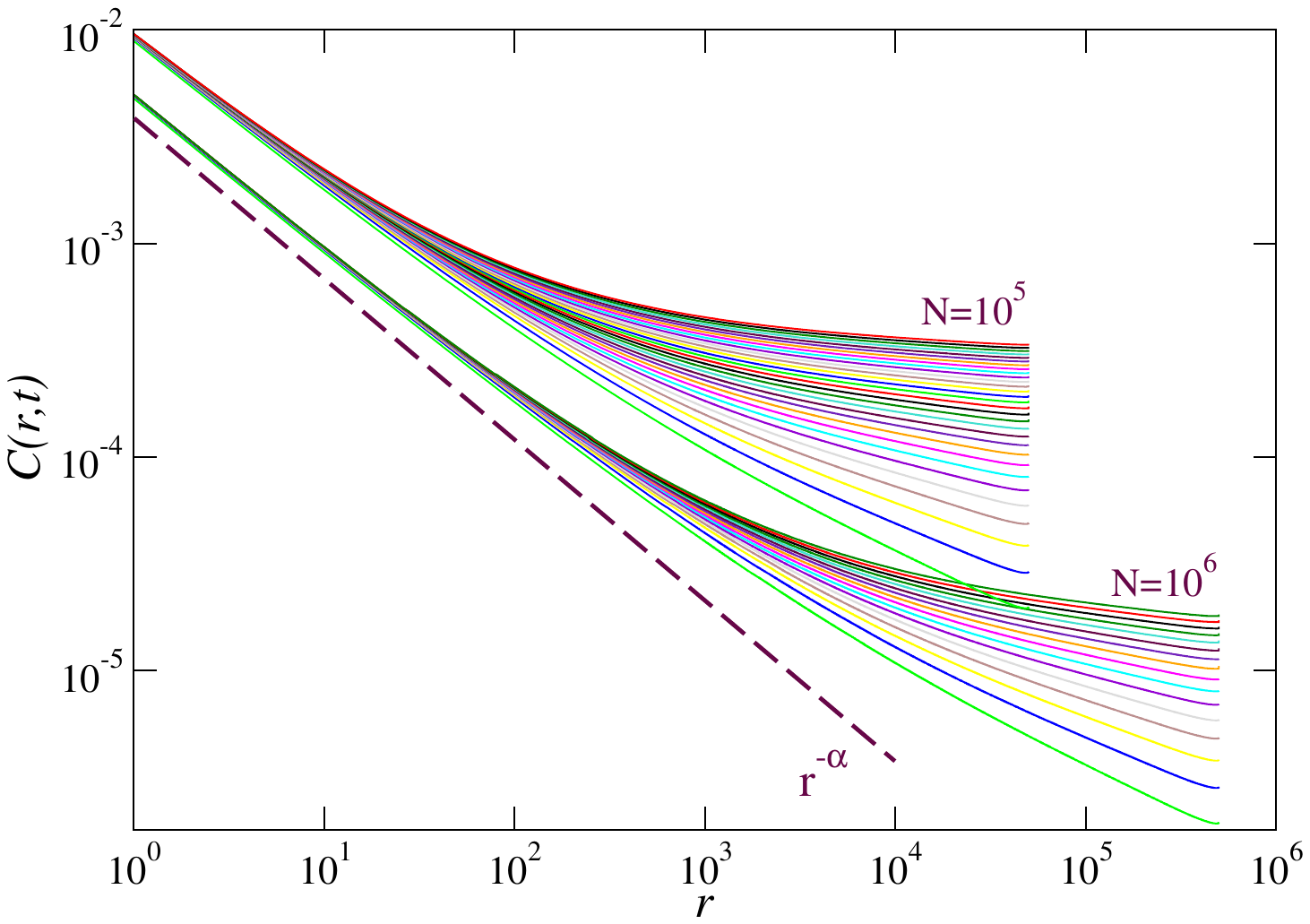}}}
	\caption{$C(r,t)$ is plotted against $r$ for $\alpha=3/2$ (left panel) and $\alpha=3/4$ (right panel). Different curves correspond to different times, increasing from bottom to top (left panel: $t=1300, 1400, \cdots$; right panel: $t=1.5, 2, 2.5, \dots$). Dashed green lines are the analytical behaviors~(\ref{Cstat},~\ref{Cstat1}). On the left panel the system size is $N=10^5$, while the two groups of curves on the right panel correspond $N=10^5$ and $N=10^6$ \cite{CorbCast24}.}
	\label{figCstat}
\end{figure}

For $\alpha <1$ we are in the SLR regime and the stationary state behaves as
\be
C_{stat}(r)=\kappa '(N)\,r^{-\alpha}\,
\label{Cstat1}
\ee
with
\be
\kappa '\propto 
N^{-(1-\alpha)},
\label{depkappa}
\ee
from which one has again a characteristic length diverging in the thermodynamic limit as:
\be
L_{stat}=\frac{1-\alpha}{2(2-\alpha)} \,N.
\label{Lstat1}
\ee
In Fig.~\ref{figCstat}, a numerical solution for $C(r,t)$ is plotted against $r$ at different times, for $\alpha=3/2$ and for
$\alpha=3/4$. It is seen that $C(r,t)$ is time-independent and obeys the analytic stationary forms~(\ref{Cstat},~\ref{Cstat1}) for sufficiently small values of $r$. The evolution observed at large values of $r$ represent the incipient escape from the stationary state because, being $N$ finite, the consensus state (i.e. $C(r)\equiv 1$) is eventually approached. 

Since $L_{stat}\propto N$ 
in the stationary state, but $L(0) \sim 1$, $L(t)$ must grow in the regime preceding stationarity, which happens by means of coarsening. In~\cite{CorbCast24}, using scaling, it is shown that
a ballistic increase 
\be
L(t)=A\,t \, 
\label{balreg}
\ee
is present.
Such growth occurs for any value of $0 \le \alpha\le 2$, however, duration of the coarsening is extensive in $N$ only for $1< \alpha \le 2$ (WLR case).

\begin{figure}[ht]
	\vspace{1.0cm}
	\centering
	\rotatebox{0}
	{\resizebox{0.75\textwidth}{!}{\includegraphics{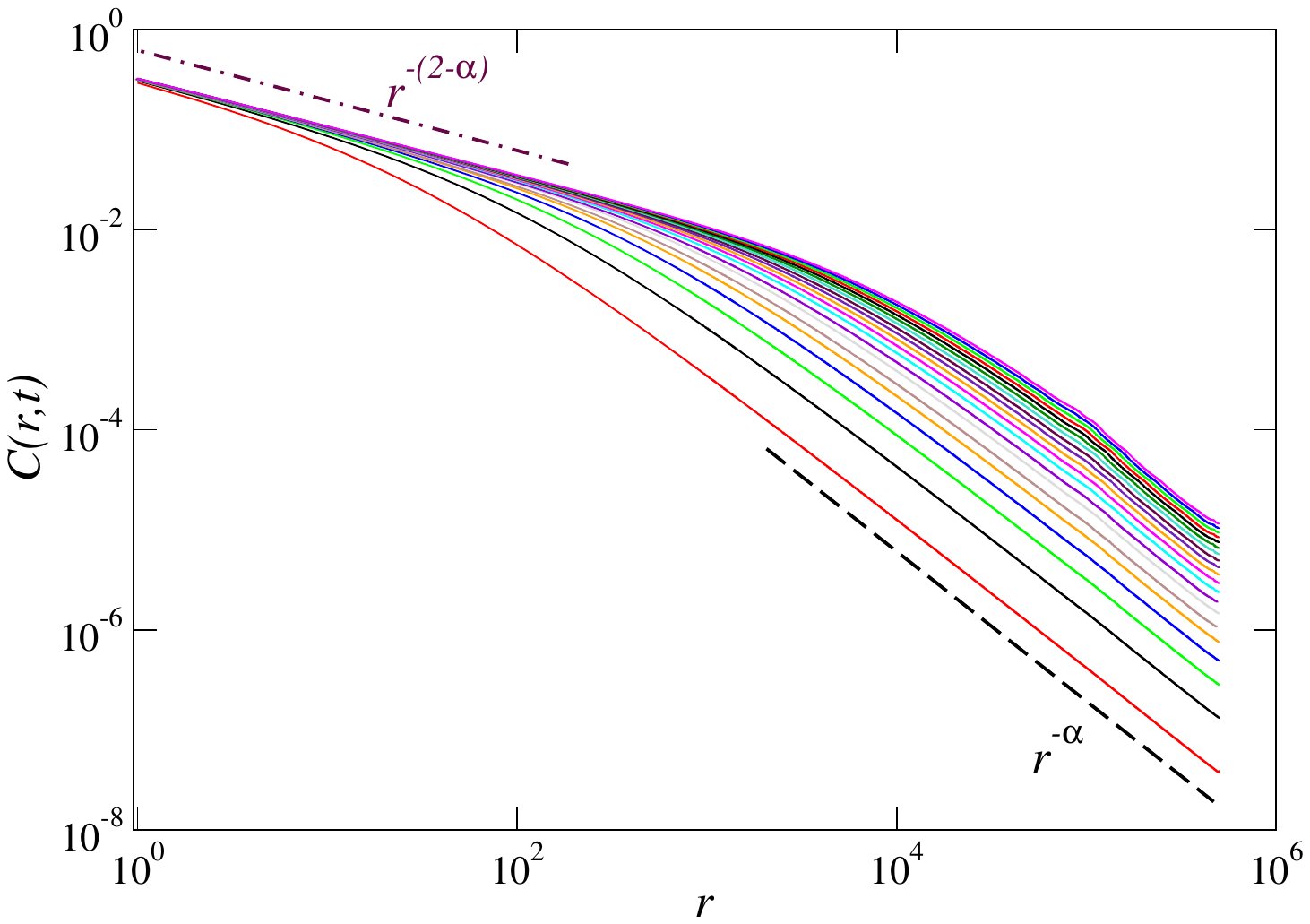}}}
	% {\resizebox{0.49\textwidth}{!}{\includegraphics{fig_scal_corr.pdf}}}
	\caption{$C(r,t)$ is plotted against $r$ %(left panel) or against $x=r/L(t)$ (right panel) 
		at different times from $t=5$ to $t=100$ in steps of $\Delta t=5$ (different colors, bottom-top) for $\alpha=3/2$. This range of times corresponds to a ballistic increase of $L(t)$. System size is $N=10^6$. The dashed blue line corresponds to the stationary form~(\ref{Cstat}), while the green line is the behavior~(\ref{unomenxbis}) \cite{CorbCast24}. %In the inset we plot $C(r,t)[1+q/L(t)]$, with $q=5\cdot10^4$, against $x$ (same data as main figure, see text for discussion).
	}
	\label{fig_scal_corr}
\end{figure}

Since the stationary state is highly correlated, it can be expected that such correlations start to develop at small distances, to proceed then to progressively larger values of $r$ as time elapses, until full stationarity.
This is clearly seen in Fig.~\ref{fig_scal_corr},
where it can be appreciated that $C(r,t)$
approaches $C_{stat}$ starting from small values of $r$.
For large $x$, the correlation function has the same form as in~(\ref{unomenxbis})

%%%%%%%%%%%%%%%%%%%%%%%%%%%%%%%%%%%%%%%%%%%%%%%%%%%%%%%%%%%%%%%%%%%%%%%%%%
\subsection{The hybrid VM and the comparison between IM and VM}

The one-dimensional hybrid VM, in which a fraction $\gamma$ of the links connect spins at distance $\ell$ via long-range interactions, while the remaining fraction $1-\gamma$ corresponds to nn interactions, has also been investigated~\cite{GLR25}. 
The results show that even a small fraction of long-range links can fundamentally alter the system's asymptotic behavior. 
For $\alpha > \alpha_{SR}=3$, their effect is limited to distances larger than the time-dependent scale $r^*(t)=x^* L(t)$, with $x^*$ defined in Sec.~\ref{agt3}. More precisely, for $r > r^*(t)$, the correlation function crosses over to a power-law decay, resembling the case of purely long-range interactions. 
In contrast, when the long-range interactions decay slowly ($\alpha < \alpha_{SR}=3$), they dominate the dynamics at all scales, leading to behavior akin to that of a fully long-range system. 
In particular, for $\alpha \le 2$, the presence of long-range links induces a stationary steady state
for arbitrarily small $\gamma$.

\section{Comparison between the IM and the VM} \label{sec_compa}

To conclude the discussion of the IM and of the VM, we present a comparison between their coarsening exponents. Since $L(t)$ and $\rho ^{-1}$, two possible definitions of the size of a growing length, may not behave similarly in the VM, as discussed before, 
we report both the exponents $a$, and $n$, relating to these two quantities, defined by 
\be
L(t)=t^a \, , 
\ee
and 
\be \label{eqRho}
\rho^{-1} (t) \sim t^n,
\ee
both valid for long times.
The value of these exponent is reported in Table~\ref{table_a}).

\begin{table}[ht]
	\centering
	\begin{tabularx}{\textwidth}{||X||c|c|c||}
		\hline \hline
		& \textbf{IM} ($n$) & \textbf{VM} ($n$) & \textbf{VM ($a$)} \\
		\hline \hline
		
		$0 \le \alpha \le \alpha_{LR} (=1)$ 
		& $1?$ 
		& $0$ 
		& $0$ \\ 
		\hline
		
		$\alpha_{LR} < \alpha \le \alpha_{SR}$ 
		& $1/\alpha$ 
		& \begin{tabular}{@{}c@{}}
			$0$ \quad {\rm if} \,   $1<\alpha \le 2$ \\[4pt]
			$(\alpha-2)/(\alpha-1)$  \quad {\rm if} \,  $\alpha>2$
		\end{tabular} 
		& \begin{tabular}{@{}c@{}}
			$1$ \quad  {\rm if} \,  $1<\alpha \le 2$ \\[4pt]
			$1/(\alpha-1)$ \quad {\rm if} \,   $\alpha>2$
		\end{tabular} \\ 
		\hline
		
		$\alpha > \alpha_{SR}$ 
		& $1/2$ 
		& $1/2$ 
		& $1/2$ \\ 
		\hline \hline
	\end{tabularx}
	
	\caption{Value of the exponents $n$ and $a$ for the IM (quenched to a sufficiently small but finite temperature) and the VM, for different ranges of $\alpha$, specified in the left column. The values of $\alpha_{SR}$ differ between IM and VM: $\alpha_{SR}(\mathrm{IM})=2$, $\alpha_{SR}(\mathrm{VM})=3$. The entry $1?$ for $0\le \alpha \le \alpha_{LR}$ in the IM indicates the putative nature of this value (see text for further discussion).
		In the IM short range, we reported the exponent $n=1/2$, which is referred to the coarsening stage before equilibration, since $T$ is finite (the duration of such stage diverges as $T \to 0$). For the VM, the exponent $a=0$ means we immediately reach a stationary state without a macroscopic coarsening in the SLR case, while $n=0$ means that the coarsening occur without reduction of the number of interfaces.}
	\label{table_a}
\end{table}

It is well known that in one dimension and with nn couplings the IM and the VM can be mapped one onto the other. This is no longer true in higher dimension or with long-range interactions. 
Moreover, as it is clear from the table, the IM and the VM with long-range interactions do not fall into the same universality class in one dimension (the same is true also in any higher dimension).
Therefore, besides some shared gross features, the VM cannot be regarded as a solvable playground to infer at a quantitative level the  dynamical properties of the IM.
Obviously, the difference between the two models springs from the fact that the update of a spin is influenced by all the others in the IM whereas it depends on the state of one single other agent in the VM.
In the next section we briefly review a model, denoted the $p$-voter model (pVM), where the number $p$ of variables influencing the attempted flip of a spin can be tuned. This model, therefore, allows, varying $p$, to move from the voter to the Ising universality class.

%%%%%%%%%%%%%%%%%%%%%%%%%%%%%%%%%%%%%%%%%%%%%%%%%%%%%%%%%%%%%%%%%%%%%%%%%%%%%%%%%%%%%%%%%%%% 

\section{Interpolating between Ising and voter: 
	\texorpdfstring{the $p$-voter model}{the p-voter model}}
\label{pVMsec}

We now discuss the p-voter model (pVM) \cite{Corberi_2024}, a generalization of the VM, where $S_i$ takes the orientation of the partial spin-average 
\be
H_i(\{S_{k_n}\}_1^p)=\frac{1}{p}\,{\sum_{n=1}^p}^* S_{k_n}
\label{eqH}
\ee
of $p$ spins $S_{k_n}$, each chosen at different distances $r_n=|k_n-i|$ 
with probability $P(r_n)$.
Since the spins are randomly selected, the same spin may appear more than once in the average, 
especially for large values of $p$.
The asterisk on the summation indicates that $k_n \neq i$ for all $n = 1, \dots, p$. The notation $\{S_{k_n}\}_1^p$ emphasizes that the function depends on the entire set of spins $S_{k_n}$ with $n = 1, \dots, p$. 

In the mean-field limit $\alpha \to 0$, this model, referred to as the $q$-Ising model, 
was originally introduced in~\cite{PhysRevE.92.052105,Chmiel2018}, where its stationary states were analyzed. It was found that a phase transition occurs only for $p > 2$, being continuous for $p = 3$ and discontinuous for $p > 3$.

The transition probability of the pVM is a generalization of Eq.(\ref{transrates}):
\be
W_{pVM}(S_i)=\frac{1}{2N}\prod _{m=1}^p \,\sum _{r_m=1}^{N/2}P(r_m)\sum _{k_m=[[i\pm r_m]]}[1-S_i \sigma (H_i)] \, ,
\label{transp}
\ee
where $\sigma(H)=\mbox{sign}(H)$ if $H\neq 0$ and $\sigma (H)=0$ when $H=0$.
Clearly, the original VM is recovered for $p=1$. Moreover, it is easy to show~\cite{Corberi_2024} that the 2VM, namely the model 
with $p=2$, can also be exactly mapped onto the VM. However, such equivalence is limited to $p=2$ where $\sigma$ is linear in $H$, which is not the case when $p>2$. 
Also the pVM, as its $p=1$ counterpart (the VM), violates detailed balance.

%%%%%%%%%%%%%%%%%%%%%%%%%%%%%%%%%%%%%%%%%%%%%%%%%%%%%%%%%%%%%%
%\subsection{\boldmath{$\alpha >1$}} \label{a>1}

\begin{figure}[ht]
	\vspace{1.0cm}
	\centering
	\rotatebox{0}{\resizebox{1.0\textwidth}{!}{\includegraphics{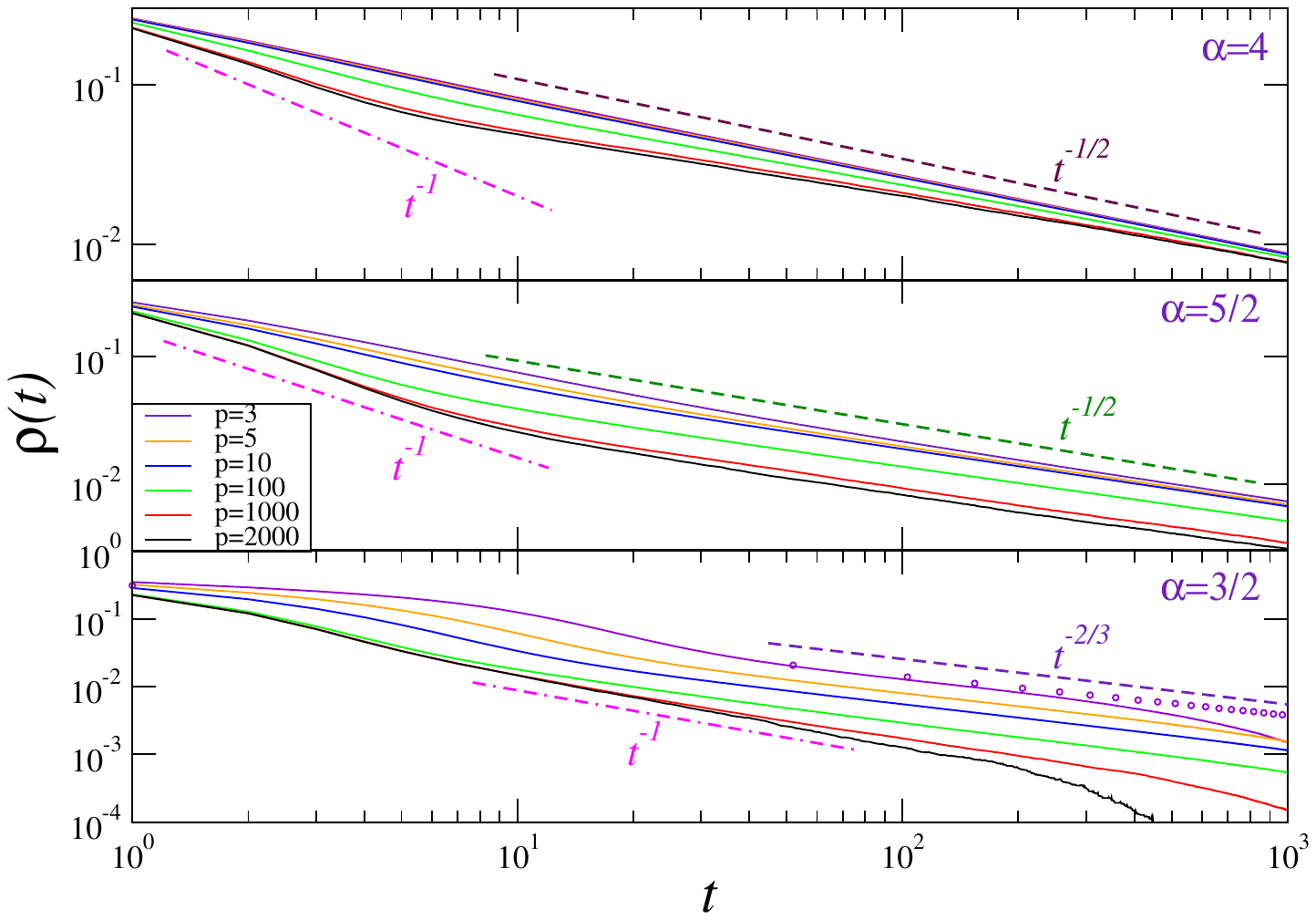}}}
	\caption{Time evolution of $\rho$, in a log-log plot, in the pVM with $p=3$, for $\alpha=4$ (top panel), $\alpha=5/2$ (central panel), and 
		$\alpha=3/2$ (bottom panel). System size is $N=10^3$ (for comparison, only for the case $\alpha=3/2$ with $p=3$, a larger system size $N=10^5$ is plotted with black open circles). A statistical average up to $5\times 10^4$ has been taken. In each panel, different curves refer to various values of $p$ (see legend).
		Dashed lines are the expected slopes $t^{-1/2}$ and $t^{-2/3}$ for the IM (at $T\neq 0$), as reported in table~\ref{table_a}.
		The dot-dashed line is the expected slope $t^{-1}$ for the IM quenched to $T= 0$) \cite{Corberi_2024}.}
	\label{figAgt1}
\end{figure}

Let us discuss the behavior of the pVM for $\al>1$: 
$\rho(t)$ is shown in Fig.~\ref{figAgt1}, with each panel corresponding to a representative value of $\alpha$ selected from the different regimes summarized in Table \ref{table_a}. Specifically, we consider: $\alpha = 4$ (with $\alpha > \alpha_{SR}$ in both the IM and the VM), $\alpha = 5/2$ (such that $2 < \alpha \le \alpha_{SR}$ for the VM but $\alpha > \alpha_{SR}$ for the IM), and $\alpha = 3/2$ (with $\alpha \le 2$ in the VM and $\alpha_{LR} \le \alpha \le \alpha_{SR}$ in the IM).

In order to assess the question of the universality class of the pVM we will concentrate on $\rho(t)$ and on its associated exponent $n$, defined in Eq.~(\ref{eqRho}). Analogous conclusions are obtained 
by considering other quantities~\cite{Corberi_2024}.
As shown in Fig. \ref{figAgt1}, regardless of the value of $p$, the asymptotic exponent $n$ typically matches that of the IM at finite temperature (see Table~\ref{table_a}): specifically, $n = 1/2$ for $\alpha = 4$ and $\alpha = 5/2$, and $n = 1/\alpha = 2/3$ for $\alpha = 3/2$. These expected behaviors are indicated by dashed lines in Fig.~\ref{figAgt1}. 

For $\alpha = 4$ and $\alpha=5/2$, the asymptotic exponent $n = 1/2$ emerges early for small $p$, whereas for larger $p$ a pre-asymptotic regime with $n \simeq 1$ is observed, which delays the onset of the final scaling behavior. 
The case $\alpha = 3/2$ is more subtle. There, finite-size effects become relevant at late times, causing a downward bending in the curves, particularly for $p = 3$ and for $p = 1000, 2000$. 
These effects can be mitigated by increasing the system size. This is demonstrated in the lower panel of Fig.~\ref{figAgt1}, where results for a larger system with $N = 10^5$ are shown for $p = 3$ (black open circles). In this case, the bending disappears and the data follow a clear algebraic decay with exponent $2/3$. These results indicate that the pVM enters the Ising universality class for $p \ge p^* \equiv 3$.

As it can be seen in Fig.~\ref{figAgt1}, besides finite-size effects, the asymptotic exponent $n$ may be obscured by a pre-asymptotic regime characterized by $n = 1$. This is particularly evident for the two largest values of $p$. Nevertheless, it is expected that the appropriate asymptotic exponent would emerge at longer times, provided simulations could extend beyond the pre-asymptotic regime without entering the finite-size dominated regime, something that would require significantly larger systems. Let us remind that $n=1$ is the exponent associated to the dynamics of the IM quenched exactly to $T=0$. 
Physically this can be explained as follows: in the pVM, when trying to flip a spin, once the $p$ spins to confront with are selected, the move is deterministic. However the choice of the $p$ spins is a stochastic event which, at least at a speculative level, can be interpreted
as a kind of temperature. This is why the model falls into the universality class of the IM quenched to finite temperatures. However, when $p\to \infty$ all the spins in the system are selected and the model becomes fully deterministic. For this reason, it falls into IM universality class at $T=0$. Clearly, for any $p$, a crossover phenomenon is observed (as  time proceeds) between the two universality classes, as indeed it is observed in Fig.~\ref{figAgt1}.

%It is worth noting that, for $\alpha > 2$, the system, similarly to the IM, eventually reaches equilibrium (corresponding to $n = 0$), but this occurs on timescales far beyond those accessible in our simulations.

Although here, for conciseness, we discussed only the case $\alpha >1$, similar conclusions are arrived at also for $\al \le 1$ where, however, the situation is enriched by the existence of different kinds of dynamical trajectories, discussed in Sec.~ \ref{simplified}.

\section{Conclusions} \label{SecConcl}

In this work, we have reviewed the main features of the one-dimensional IM with long-range 
interactions, characterized by spin--spin couplings decaying algebraically as 
$J(r) \propto r^{-\alpha}$. Alongside this, we discussed the properties of the 
one-dimensional VM, in which two agents (spins) interact with a similar 
power-law probability. A comparative analysis of the two models was presented, together 
with the pVM, which serves as an interpolation scheme between them. In fact, 
the pVM reduces to the VM for $p=1$ and $p=2$, while for $p \geq 3$ 
it falls into the universality class of the IM at finite $T$ below $T_c$. Notably, in the limiting case 
$p \to \infty$, the model reproduces the behavior of the zero-temperature IM.

An interesting case we have not reviewed in this paper is the one of conserved order parameter (COP) dynamics for the IM \cite{CLP_review}, where the basic process is the exchange between opposite spins $S_i \leftrightarrow S_j$ (Kawasaki dynamics). This can also be described, defining $2 n_i=1+S_i$ ($n_i=0,1$), as the dynamics of a lattice gas, where $n_i=1$ represents a particle and $n_i=0$ a hole. Even in that case, one finds that the asymptotic growth law for $L(t)$ in the coarsening dynamics is the usual algebraic one, $L(t)\sim t^a$, 
of the corresponding model with nn interaction (i.e. $a=1/3$) for all values $\al>1$. For $\al \to 1^+$ one finds $a=1/(4\beta+3)$. 
From this last expression is evident that for COP dynamics temperature plays a more relevant role than in the non-conserved case we have reviewed in this paper.

Aging properties of long-range spin models, which are encoded into the 
two-time correlation function $G(r;t,s)$ ($t\ge s$) between the state of a variable $S_i$ at time $s$ and that of another one, at distance $r$, at time $t$, is another aspect that we have not reviewed in this work. Of particular interest is the autocorrelation function $A(t,s)=G(r=0;t,s)$. This usually decays algebraically for $\alpha >1$ as $[L(t)/L(s)]^{-\lambda}$, where $\lambda $ is the Fisher-Huse exponent.
In both IM \cite{Corberi_2019} and VM model~\cite{corberi2024aging}, in $D=1$, the value $\alpha =2$ separates a region $\alpha >2$ where the autocorrelation exponent conserves the nn value~\cite{Glauber} $\lambda =1$
from the region $\alpha \le 2$ where a different value, due to
the long-range character, is observed. This small-$\alpha $ value is 
$\lambda =1/2$ for Ising, whereas it is $\lambda =1/(\alpha -1)$
for voter. For $\alpha \le 1$ (SLR regime) there is an exponential decay, as in the mean field case.

Although this article was devoted to one-dimensional systems, which are better understood, the behavior of spin models with long-range interactions in $D>1$ is also, of course, of paramount interest. 
Let us also briefly comment on 
this. 

Concerning the $D=2$ long-range IM \cite{PhysRevE.103.012108}, for quenches to any nonzero temperature $T < T_c$, 
the domain size still follows the expression $L(t)=t^{a}$, with dynamic exponent $a = 1/(\al-1)$ for $\al < 3$, 
crossing over to the diffusive value $a=1/2$ for $\al > 3$. In contrast, at strictly zero 
temperature, long-range interactions induce a persistent drift of domain interfaces, which 
qualitatively modifies the coarsening kinetics. In this case, the growth exponent takes on 
the universal value $a=3/4$, independent of $\al$.

In the VM in $D$ dimensions, one still identifies a value $\alpha_{SR}=D+2$, such that for $\alpha>\alpha_{SR}$ the behavior resembles that of the nn case. Partially ordered stationary states exist for any $\alpha$ only when $D\geq 3$. In particular, for $D\geq 3$ and $\alpha>\alpha_{SR}$, the stationary state is characterized by the correlation function $C_{stat}(r)\propto r^{2-D}$, as in the nn case~\cite{krapivsky2010kinetic}.
For $\alpha \leq \alpha_{SR}$, stationary states without consensus also appear in lower dimensions. More precisely, such states occur for $\alpha \leq 4$ in $D=2$, and for $\alpha \leq 2$ in $D=1$ (the latter case being reviewed in this work). The functional form of $C_{stat}$ depends on whether $\alpha_{LR}<\alpha \leq \alpha_{SR}$ or $0\leq \alpha \leq \alpha_{LR}$, where $\alpha_{LR}=D$ marks the value below which $P(r)$ becomes non-summable.
For $\alpha_{LR}<\alpha \leq \alpha_{SR}$, one finds $C_{stat}(r)\propto r^{-(2D-\alpha)}$. Conversely, for $0\leq \alpha \leq \alpha_{LR}$, one has $C_{stat}(r)\propto r^{-\alpha}$ \cite{CorbCast24,corsmal2023ordering,corberi20243d}, which we also expect to be valid in arbitrary dimensions. In all cases, the forms of $C_{stat}$ together with the values of $\alpha_{SR}$ and $\alpha_{LR}$ imply that $C_{stat}$ is always a decreasing, non-summable function. %Since the decay exponent %$C_{stat}(r)\sim r^{-\gamma}$ is %related to the fractal dimension of %the structure through %$D_f=D-\frac{\gamma}{2}$ %\cite{CONIGLIO2000129},
%we obtain $D_f = \tfrac{D}{2}+1$, %$D_f = \tfrac{\alpha}{2}$, or $D_f = %D-\tfrac{\alpha}{2}$ in the three %regimes discussed above, namely %$\alpha>\alpha_{SR}$ (valid only for %$D\geq 3$), $\alpha_{LR}<\alpha\leq %\alpha_{SR}$, and $0\leq \alpha \leq %\alpha_{LR}$,

All in all, the above discussion shows the very rich phenomenology induced by the presence of long-range interactions in the kinetic properties of spin models driven out of equilibrium. Nowadays, as we have reviewed in this paper, understanding of the behavior of such systems is starting to be understood. Many topics, however, remain to be clarified, particularly in the SLR regime and in $D>1$.  
The topic, therefore, still remains a challenging playground to achieve a better understanding of long-range systems in general.

\section*{Acknowledgments}

FC and EL acknowledge financial support by MUR PRIN PNRR 2022B3WCFS.
PP acknowledges support from the MUR PRIN2022 project ‘Breakdown of
ergodicity in classical and quantum many-body systems’ (BECQuMB) Grant No.
20222BHC9Z.

	\bibliography{LibraryStat}   % name your BibTeX data base

%% BioMed_Central_Bib_Style_v1.01

\begin{thebibliography}{56}
% BibTex style file: bmc-mathphys.bst (version 2.1), 2014-07-24
\ifx \bisbn   \undefined \def \bisbn  #1{ISBN #1}\fi
\ifx \binits  \undefined \def \binits#1{#1}\fi
\ifx \bauthor  \undefined \def \bauthor#1{#1}\fi
\ifx \batitle  \undefined \def \batitle#1{#1}\fi
\ifx \bjtitle  \undefined \def \bjtitle#1{#1}\fi
\ifx \bvolume  \undefined \def \bvolume#1{\textbf{#1}}\fi
\ifx \byear  \undefined \def \byear#1{#1}\fi
\ifx \bissue  \undefined \def \bissue#1{#1}\fi
\ifx \bfpage  \undefined \def \bfpage#1{#1}\fi
\ifx \blpage  \undefined \def \blpage #1{#1}\fi
\ifx \burl  \undefined \def \burl#1{\textsf{#1}}\fi
\ifx \doiurl  \undefined \def \doiurl#1{\url{https://doi.org/#1}}\fi
\ifx \betal  \undefined \def \betal{\textit{et al.}}\fi
\ifx \binstitute  \undefined \def \binstitute#1{#1}\fi
\ifx \binstitutionaled  \undefined \def \binstitutionaled#1{#1}\fi
\ifx \bctitle  \undefined \def \bctitle#1{#1}\fi
\ifx \beditor  \undefined \def \beditor#1{#1}\fi
\ifx \bpublisher  \undefined \def \bpublisher#1{#1}\fi
\ifx \bbtitle  \undefined \def \bbtitle#1{#1}\fi
\ifx \bedition  \undefined \def \bedition#1{#1}\fi
\ifx \bseriesno  \undefined \def \bseriesno#1{#1}\fi
\ifx \blocation  \undefined \def \blocation#1{#1}\fi
\ifx \bsertitle  \undefined \def \bsertitle#1{#1}\fi
\ifx \bsnm \undefined \def \bsnm#1{#1}\fi
\ifx \bsuffix \undefined \def \bsuffix#1{#1}\fi
\ifx \bparticle \undefined \def \bparticle#1{#1}\fi
\ifx \barticle \undefined \def \barticle#1{#1}\fi
\bibcommenthead
\ifx \bconfdate \undefined \def \bconfdate #1{#1}\fi
\ifx \botherref \undefined \def \botherref #1{#1}\fi
\ifx \url \undefined \def \url#1{\textsf{#1}}\fi
\ifx \bchapter \undefined \def \bchapter#1{#1}\fi
\ifx \bbook \undefined \def \bbook#1{#1}\fi
\ifx \bcomment \undefined \def \bcomment#1{#1}\fi
\ifx \oauthor \undefined \def \oauthor#1{#1}\fi
\ifx \citeauthoryear \undefined \def \citeauthoryear#1{#1}\fi
\ifx \endbibitem  \undefined \def \endbibitem {}\fi
\ifx \bconflocation  \undefined \def \bconflocation#1{#1}\fi
\ifx \arxivurl  \undefined \def \arxivurl#1{\textsf{#1}}\fi
\csname PreBibitemsHook\endcsname

%%% 1
\bibitem[\protect\citeauthoryear{Huang}{1987}]{Huang1987}
\begin{bbook}
\bauthor{\bsnm{Huang}, \binits{K.}}:
\bbtitle{Statistical Mechanics},
\bedition{2}nd edn.
\bpublisher{John Wiley \& Sons},
\blocation{New York}
(\byear{1987})
\end{bbook}
\endbibitem

%%% 2
\bibitem[\protect\citeauthoryear{Pathria and Beale}{2011}]{Pathria2011}
\begin{bbook}
\bauthor{\bsnm{Pathria}, \binits{R.K.}},
\bauthor{\bsnm{Beale}, \binits{P.D.}}:
\bbtitle{Statistical Mechanics},
\bedition{3}rd edn.
\bpublisher{Elsevier},
\blocation{Amsterdam}
(\byear{2011})
\end{bbook}
\endbibitem

%%% 3
\bibitem[\protect\citeauthoryear{Baxter}{1982}]{Baxter1982}
\begin{bbook}
\bauthor{\bsnm{Baxter}, \binits{R.J.}}:
\bbtitle{Exactly Solved Models in Statistical Mechanics}.
\bpublisher{Academic Press},
\blocation{London}
(\byear{1982})
\end{bbook}
\endbibitem

%%% 4
\bibitem[\protect\citeauthoryear{Krapivsky et~al.}{2010}]{krapivsky2010kinetic}
\begin{bbook}
\bauthor{\bsnm{Krapivsky}, \binits{P.L.}},
\bauthor{\bsnm{Redner}, \binits{S.}},
\bauthor{\bsnm{Ben-Naim}, \binits{E.}}:
\bbtitle{A Kinetic View of Statistical Physics}.
\bpublisher{Cambridge University Press}, \blocation{???}
(\byear{2010}).
\burl{https://books.google.it/books?id=cc3pApnX3kYC}
\end{bbook}
\endbibitem

%%% 5
\bibitem[\protect\citeauthoryear{Livi and Politi}{2017}]{LiviPoliti2017}
\begin{bbook}
\bauthor{\bsnm{Livi}, \binits{R.}},
\bauthor{\bsnm{Politi}, \binits{P.}}:
\bbtitle{Nonequilibrium Statistical Physics: A Modern Perspective}.
\bpublisher{Cambridge University Press},
\blocation{Cambridge}
(\byear{2017})
\end{bbook}
\endbibitem

%%% 6
\bibitem[\protect\citeauthoryear{Bray and Rutenberg}{1994}]{BrayRut94}
\begin{barticle}
\bauthor{\bsnm{Bray}, \binits{A.J.}},
\bauthor{\bsnm{Rutenberg}, \binits{A.D.}}:
\batitle{Growth laws for phase ordering}.
\bjtitle{Phys. Rev. E}
\bvolume{49},
\bfpage{27}--\blpage{30}
(\byear{1994})
\doiurl{10.1103/PhysRevE.49.R27}
\end{barticle}
\endbibitem

%%% 7
\bibitem[\protect\citeauthoryear{Rutenberg and Bray}{1994}]{RutBray94}
\begin{barticle}
\bauthor{\bsnm{Rutenberg}, \binits{A.D.}},
\bauthor{\bsnm{Bray}, \binits{A.J.}}:
\batitle{Phase-ordering kinetics of one-dimensional nonconserved scalar
  systems}.
\bjtitle{Phys. Rev. E}
\bvolume{50},
\bfpage{1900}--\blpage{1911}
(\byear{1994})
\doiurl{10.1103/PhysRevE.50.1900}
\end{barticle}
\endbibitem

%%% 8
\bibitem[\protect\citeauthoryear{Christiansen et~al.}{2019}]{CMJ19}
\begin{barticle}
\bauthor{\bsnm{Christiansen}, \binits{H.}},
\bauthor{\bsnm{Majumder}, \binits{S.}},
\bauthor{\bsnm{Janke}, \binits{W.}}:
\batitle{Phase ordering kinetics of the long-range ising model}.
\bjtitle{Phys. Rev. E}
\bvolume{99},
\bfpage{011301}
(\byear{2019})
\doiurl{10.1103/PhysRevE.99.011301}
\end{barticle}
\endbibitem

%%% 9
\bibitem[\protect\citeauthoryear{Corberi et~al.}{2017}]{Corberi_2017}
\begin{barticle}
\bauthor{\bsnm{Corberi}, \binits{F.}},
\bauthor{\bsnm{Lippiello}, \binits{E.}},
\bauthor{\bsnm{Politi}, \binits{P.}}:
\batitle{Effective mobility and diffusivity in coarsening processes}.
\bjtitle{Europhysics Letters}
\bvolume{119}(\bissue{2}),
\bfpage{26005}
(\byear{2017})
\doiurl{10.1209/0295-5075/119/26005}
\end{barticle}
\endbibitem

%%% 10
\bibitem[\protect\citeauthoryear{{Corberi} et~al.}{2019}]{Corberi2019JSM}
\begin{barticle}
\bauthor{\bsnm{{Corberi}}, \binits{F.}},
\bauthor{\bsnm{{Cugliandolo}}, \binits{L.F.}},
\bauthor{\bsnm{{Insalata}}, \binits{F.}},
\bauthor{\bsnm{{Picco}}, \binits{M.}}:
\batitle{{Fractal character of the phase ordering kinetics of a diluted
  ferromagnet}}.
\bjtitle{Journal of Statistical Mechanics: Theory and Experiment}
\bvolume{4}(\bissue{4}),
\bfpage{043203}
(\byear{2019})
\doiurl{10.1088/1742-5468/ab02ee}
{\href{https://arxiv.org/abs/1811.12675}{{arXiv:1811.12675}}}
{[cond-mat.stat-mech]}
\end{barticle}
\endbibitem

%%% 11
\bibitem[\protect\citeauthoryear{Agrawal et~al.}{2021}]{PhysRevE.103.012108}
\begin{barticle}
\bauthor{\bsnm{Agrawal}, \binits{R.}},
\bauthor{\bsnm{Corberi}, \binits{F.}},
\bauthor{\bsnm{Lippiello}, \binits{E.}},
\bauthor{\bsnm{Politi}, \binits{P.}},
\bauthor{\bsnm{Puri}, \binits{S.}}:
\batitle{Kinetics of the two-dimensional long-range ising model at low
  temperatures}.
\bjtitle{Phys. Rev. E}
\bvolume{103},
\bfpage{012108}
(\byear{2021})
\doiurl{10.1103/PhysRevE.103.012108}
\end{barticle}
\endbibitem

%%% 12
\bibitem[\protect\citeauthoryear{{Corberi} et~al.}{2021}]{Corberi2021SCI}
\begin{barticle}
\bauthor{\bsnm{{Corberi}}, \binits{F.}},
\bauthor{\bsnm{{Iannone}}, \binits{A.}},
\bauthor{\bsnm{{Kumar}}, \binits{M.}},
\bauthor{\bsnm{{Lippiello}}, \binits{E.}},
\bauthor{\bsnm{{Politi}}, \binits{P.}}:
\batitle{{Coexistence of coarsening and mean field relaxation in the long-range
  Ising chain}}.
\bjtitle{SciPost Physics}
\bvolume{10}(\bissue{5}),
\bfpage{109}
(\byear{2021})
\doiurl{10.21468/SciPostPhys.10.5.109}
{\href{https://arxiv.org/abs/2102.08217}{{arXiv:2102.08217}}}
{[cond-mat.stat-mech]}
\end{barticle}
\endbibitem

%%% 13
\bibitem[\protect\citeauthoryear{Corberi et~al.}{2023}]{CORBERI2023113681}
\begin{barticle}
\bauthor{\bsnm{Corberi}, \binits{F.}},
\bauthor{\bsnm{Kumar}, \binits{M.}},
\bauthor{\bsnm{Lippiello}, \binits{E.}},
\bauthor{\bsnm{Politi}, \binits{P.}}:
\batitle{Domain statistics in the relaxation of the one-dimensional ising model
  with strong long-range interactions}.
\bjtitle{Chaos, Solitons \& Fractals}
\bvolume{173},
\bfpage{113681}
(\byear{2023})
\doiurl{10.1016/j.chaos.2023.113681}
\end{barticle}
\endbibitem

%%% 14
\bibitem[\protect\citeauthoryear{{Agrawal} et~al.}{2023}]{Corberi2023PRE}
\begin{barticle}
\bauthor{\bsnm{{Agrawal}}, \binits{R.}},
\bauthor{\bsnm{{Corberi}}, \binits{F.}},
\bauthor{\bsnm{{Lippiello}}, \binits{E.}},
\bauthor{\bsnm{{Puri}}, \binits{S.}}:
\batitle{{Phase ordering dynamics of the random-field long-range Ising model in
  one dimension}}.
\bjtitle{Phys. Rev. E}
\bvolume{108}(\bissue{4}),
\bfpage{044131}
(\byear{2023})
\doiurl{10.1103/PhysRevE.108.044131}
{\href{https://arxiv.org/abs/2305.06723}{{arXiv:2305.06723}}}
{[cond-mat.stat-mech]}
\end{barticle}
\endbibitem

%%% 15
\bibitem[\protect\citeauthoryear{Corberi et~al.}{2020}]{PhysRevE.102.020102}
\begin{barticle}
\bauthor{\bsnm{Corberi}, \binits{F.}},
\bauthor{\bsnm{Lippiello}, \binits{E.}},
\bauthor{\bsnm{Politi}, \binits{P.}}:
\batitle{Quasideterministic dynamics, memory effects, and lack of
  self-averaging in the relaxation of quenched ferromagnets}.
\bjtitle{Phys. Rev. E}
\bvolume{102},
\bfpage{020102}
(\byear{2020})
\doiurl{10.1103/PhysRevE.102.020102}
\end{barticle}
\endbibitem

%%% 16
\bibitem[\protect\citeauthoryear{Corberi and Castellano}{2024}]{CorbCast24}
\begin{barticle}
\bauthor{\bsnm{Corberi}, \binits{F.}},
\bauthor{\bsnm{Castellano}, \binits{C.}}:
\batitle{Kinetics of the one-dimensional voter model with long-range
  interactions}.
\bjtitle{Journal of Physics: Complexity}
\bvolume{5}(\bissue{2}),
\bfpage{025021}
(\byear{2024})
\doiurl{10.1088/2632-072X/ad4dfb}
\end{barticle}
\endbibitem

%%% 17
\bibitem[\protect\citeauthoryear{Corberi and
  Smaldone}{2024a}]{corsmal2023ordering}
\begin{barticle}
\bauthor{\bsnm{Corberi}, \binits{F.}},
\bauthor{\bsnm{Smaldone}, \binits{L.}}:
\batitle{Ordering kinetics of the two-dimensional voter model with long-range
  interactions}.
\bjtitle{Phys. Rev. E}
\bvolume{109},
\bfpage{034133}
(\byear{2024})
\doiurl{10.1103/PhysRevE.109.034133}
\end{barticle}
\endbibitem

%%% 18
\bibitem[\protect\citeauthoryear{Corberi and
  Smaldone}{2024b}]{corberi2024aging}
\begin{barticle}
\bauthor{\bsnm{Corberi}, \binits{F.}},
\bauthor{\bsnm{Smaldone}, \binits{L.}}:
\batitle{Aging properties of the voter model with long-range interactions}.
\bjtitle{Journal of Statistical Mechanics: Theory and Experiment}
\bvolume{2024}(\bissue{5}),
\bfpage{053204}
(\byear{2024})
\doiurl{10.1088/1742-5468/ad41db}
\end{barticle}
\endbibitem

%%% 19
\bibitem[\protect\citeauthoryear{Corberi et~al.}{2024}]{corberi20243d}
\begin{barticle}
\bauthor{\bsnm{Corberi}, \binits{F.}},
\bauthor{\bsnm{Russo}, \binits{S.}},
\bauthor{\bsnm{Smaldone}, \binits{L.}}:
\batitle{Coarsening and metastability of the long-range voter model in three
  dimensions}.
\bjtitle{Phys. Rev. E}
\bvolume{110},
\bfpage{024143}
(\byear{2024})
\doiurl{10.1103/PhysRevE.110.024143}
\end{barticle}
\endbibitem

%%% 20
\bibitem[\protect\citeauthoryear{Bray}{1994}]{Bray94}
\begin{barticle}
\bauthor{\bsnm{Bray}, \binits{A.J.}}:
\batitle{Theory of phase-ordering kinetics}.
\bjtitle{Advances in Physics}
\bvolume{43}(\bissue{3}),
\bfpage{357}--\blpage{459}
(\byear{1994})
{\href{https://arxiv.org/abs/https://doi.org/10.1080/00018739400101505}{{https://doi.org/10.1080/00018739400101505}}}
\end{barticle}
\endbibitem

%%% 21
\bibitem[\protect\citeauthoryear{Lenz}{1920}]{Lenz1920}
\begin{barticle}
\bauthor{\bsnm{Lenz}, \binits{W.}}:
\batitle{{Beitr{\"a}ge zum Verst{\"a}ndnis der magnetischen Eigenschaften in
  festen K{\"o}rpern}}.
\bjtitle{Physikalische Zeitschrift}
\bvolume{21},
\bfpage{613}--\blpage{615}
(\byear{1920})
\end{barticle}
\endbibitem

%%% 22
\bibitem[\protect\citeauthoryear{Ising}{1925}]{Ising1925}
\begin{barticle}
\bauthor{\bsnm{Ising}, \binits{E.}}:
\batitle{Beitrag zur theorie des ferromagnetismus}.
\bjtitle{Zeitschrift f{\"u}r Physik}
\bvolume{31},
\bfpage{253}--\blpage{258}
(\byear{1925})
\doiurl{10.1007/BF02980577}
\end{barticle}
\endbibitem

%%% 23
\bibitem[\protect\citeauthoryear{Onsager}{1944}]{PhysRev.65.117}
\begin{barticle}
\bauthor{\bsnm{Onsager}, \binits{L.}}:
\batitle{Crystal statistics. i. a two-dimensional model with an order-disorder
  transition}.
\bjtitle{Phys. Rev.}
\bvolume{65},
\bfpage{117}--\blpage{149}
(\byear{1944})
\doiurl{10.1103/PhysRev.65.117}
\end{barticle}
\endbibitem

%%% 24
\bibitem[\protect\citeauthoryear{BRUSH}{1967}]{RevModPhys.39.883}
\begin{barticle}
\bauthor{\bsnm{BRUSH}, \binits{S.G.}}:
\batitle{History of the lenz-ising model}.
\bjtitle{Rev. Mod. Phys.}
\bvolume{39},
\bfpage{883}--\blpage{893}
(\byear{1967})
\doiurl{10.1103/RevModPhys.39.883}
\end{barticle}
\endbibitem

%%% 25
\bibitem[\protect\citeauthoryear{Glauber}{1963}]{Glauber}
\begin{botherref}
\oauthor{\bsnm{Glauber}, \binits{R.J.}}:
Time-dependent statistics of the ising model.
Journal of Mathematical Physics (New York) (U.S.)
\textbf{Vol: 4}
(1963)
\end{botherref}
\endbibitem

%%% 26
\bibitem[\protect\citeauthoryear{Corberi et~al.}{2019a}]{CLP_review}
\begin{barticle}
\bauthor{\bsnm{Corberi}, \binits{F.}},
\bauthor{\bsnm{Lippiello}, \binits{E.}},
\bauthor{\bsnm{Politi}, \binits{P.}}:
\batitle{One dimensional phase-ordering in the ising model with space decaying
  interactions}.
\bjtitle{Journal of Statistical Physics}
\bvolume{176}(\bissue{3}),
\bfpage{510}--\blpage{540}
(\byear{2019})
\end{barticle}
\endbibitem

%%% 27
\bibitem[\protect\citeauthoryear{Corberi et~al.}{2019b}]{Corberi_2019}
\begin{barticle}
\bauthor{\bsnm{Corberi}, \binits{F.}},
\bauthor{\bsnm{Lippiello}, \binits{E.}},
\bauthor{\bsnm{Politi}, \binits{P.}}:
\batitle{Universality in the time correlations of the long-range 1d ising
  model}.
\bjtitle{Journal of Statistical Mechanics: Theory and Experiment}
\bvolume{2019}(\bissue{7}),
\bfpage{074002}
(\byear{2019})
\doiurl{10.1088/1742-5468/ab270a}
\end{barticle}
\endbibitem

%%% 28
\bibitem[\protect\citeauthoryear{Clifford and Sudbury}{1973}]{Clifford1973}
\begin{barticle}
\bauthor{\bsnm{Clifford}, \binits{P.}},
\bauthor{\bsnm{Sudbury}, \binits{A.}}:
\batitle{A model for spatial conflict}.
\bjtitle{Biometrika}
\bvolume{60}(\bissue{3}),
\bfpage{581}--\blpage{588}
(\byear{1973})
\doiurl{10.1093/biomet/60.3.581}
\end{barticle}
\endbibitem

%%% 29
\bibitem[\protect\citeauthoryear{Holley and Liggett}{1975}]{Holley1975}
\begin{barticle}
\bauthor{\bsnm{Holley}, \binits{R.A.}},
\bauthor{\bsnm{Liggett}, \binits{T.M.}}:
\batitle{Ergodic theorems for weakly interacting infinite systems and the voter
  model}.
\bjtitle{The Annals of Probability}
\bvolume{3}(\bissue{4}),
\bfpage{643}--\blpage{663}
(\byear{1975})
\doiurl{10.2307/2959329}
\end{barticle}
\endbibitem

%%% 30
\bibitem[\protect\citeauthoryear{Liggett}{2004}]{liggett2004interacting}
\begin{bbook}
\bauthor{\bsnm{Liggett}, \binits{T.M.}}:
\bbtitle{Interacting Particle Systems}.
\bsertitle{Classics in Mathematics}.
\bpublisher{Springer}, \blocation{???}
(\byear{2004}).
\burl{https://books.google.it/books?id=I3aNPR1FursC}
\end{bbook}
\endbibitem

%%% 31
\bibitem[\protect\citeauthoryear{Cox and Griffeath}{1986}]{Theodore1986}
\begin{barticle}
\bauthor{\bsnm{Cox}, \binits{J.T.}},
\bauthor{\bsnm{Griffeath}, \binits{D.}}:
\batitle{Diffusive clustering in the two dimensional voter model}.
\bjtitle{The Annals of Probability}
\bvolume{14}(\bissue{2}),
\bfpage{347}--\blpage{370}
(\byear{1986}).
Accessed 2023-11-10
\end{barticle}
\endbibitem

%%% 32
\bibitem[\protect\citeauthoryear{{Scheucher} and {Spohn}}{1988}]{Scheucher1988}
\begin{barticle}
\bauthor{\bsnm{{Scheucher}}, \binits{M.}},
\bauthor{\bsnm{{Spohn}}, \binits{H.}}:
\batitle{{A soluble kinetic model for spinodal decomposition}}.
\bjtitle{Journal of Statistical Physics}
\bvolume{53}(\bissue{1-2}),
\bfpage{279}--\blpage{294}
(\byear{1988})
\doiurl{10.1007/BF01011557}
\end{barticle}
\endbibitem

%%% 33
\bibitem[\protect\citeauthoryear{Krapivsky}{1992}]{PhysRevA.45.1067}
\begin{barticle}
\bauthor{\bsnm{Krapivsky}, \binits{P.L.}}:
\batitle{Kinetics of monomer-monomer surface catalytic reactions}.
\bjtitle{Phys. Rev. A}
\bvolume{45},
\bfpage{1067}--\blpage{1072}
(\byear{1992})
\doiurl{10.1103/PhysRevA.45.1067}
\end{barticle}
\endbibitem

%%% 34
\bibitem[\protect\citeauthoryear{Frachebourg and
  Krapivsky}{1996}]{Frachebourg1996}
\begin{barticle}
\bauthor{\bsnm{Frachebourg}, \binits{L.}},
\bauthor{\bsnm{Krapivsky}, \binits{P.L.}}:
\batitle{Exact results for kinetics of catalytic reactions}.
\bjtitle{Phys. Rev. E}
\bvolume{53},
\bfpage{3009}--\blpage{3012}
(\byear{1996})
\doiurl{10.1103/PhysRevE.53.R3009}
\end{barticle}
\endbibitem

%%% 35
\bibitem[\protect\citeauthoryear{Ben-Naim et~al.}{1996}]{Ben1996}
\begin{barticle}
\bauthor{\bsnm{Ben-Naim}, \binits{E.}},
\bauthor{\bsnm{Frachebourg}, \binits{L.}},
\bauthor{\bsnm{Krapivsky}, \binits{P.L.}}:
\batitle{Coarsening and persistence in the voter model}.
\bjtitle{Phys. Rev. E}
\bvolume{53},
\bfpage{3078}--\blpage{3087}
(\byear{1996})
\doiurl{10.1103/PhysRevE.53.3078}
\end{barticle}
\endbibitem

%%% 36
\bibitem[\protect\citeauthoryear{Sood and Redner}{2005}]{PhysRevLett.94.178701}
\begin{barticle}
\bauthor{\bsnm{Sood}, \binits{V.}},
\bauthor{\bsnm{Redner}, \binits{S.}}:
\batitle{Voter model on heterogeneous graphs}.
\bjtitle{Phys. Rev. Lett.}
\bvolume{94},
\bfpage{178701}
(\byear{2005})
\doiurl{10.1103/PhysRevLett.94.178701}
\end{barticle}
\endbibitem

%%% 37
\bibitem[\protect\citeauthoryear{Sood et~al.}{2008}]{PhysRevE.77.041121}
\begin{barticle}
\bauthor{\bsnm{Sood}, \binits{V.}},
\bauthor{\bsnm{Antal}, \binits{T.}},
\bauthor{\bsnm{Redner}, \binits{S.}}:
\batitle{Voter models on heterogeneous networks}.
\bjtitle{Phys. Rev. E}
\bvolume{77},
\bfpage{041121}
(\byear{2008})
\doiurl{10.1103/PhysRevE.77.041121}
\end{barticle}
\endbibitem

%%% 38
\bibitem[\protect\citeauthoryear{Castellano et~al.}{2009}]{Castellano09}
\begin{barticle}
\bauthor{\bsnm{Castellano}, \binits{C.}},
\bauthor{\bsnm{Fortunato}, \binits{S.}},
\bauthor{\bsnm{Loreto}, \binits{V.}}:
\batitle{Statistical physics of social dynamics}.
\bjtitle{Rev. Mod. Phys.}
\bvolume{81},
\bfpage{591}--\blpage{646}
(\byear{2009})
\doiurl{10.1103/RevModPhys.81.591}
\end{barticle}
\endbibitem

%%% 39
\bibitem[\protect\citeauthoryear{de~Almeida et~al.}{2025}]{yjf2-4z1d}
\begin{barticle}
\bauthor{\bsnm{Almeida}, \binits{R.G.}},
\bauthor{\bsnm{Arenzon}, \binits{J.J.}},
\bauthor{\bsnm{Corberi}, \binits{F.}},
\bauthor{\bsnm{Dantas}, \binits{W.G.}},
\bauthor{\bsnm{Smaldone}, \binits{L.}}:
\batitle{Coarsening in the persistent voter model: Analytical results}.
\bjtitle{Phys. Rev. E}
\bvolume{111},
\bfpage{064313}
(\byear{2025})
\doiurl{10.1103/yjf2-4z1d}
\end{barticle}
\endbibitem

%%% 40
\bibitem[\protect\citeauthoryear{Zillio et~al.}{2005}]{Zillio2005}
\begin{barticle}
\bauthor{\bsnm{Zillio}, \binits{T.}},
\bauthor{\bsnm{Volkov}, \binits{I.}},
\bauthor{\bsnm{Banavar}, \binits{J.R.}},
\bauthor{\bsnm{Hubbell}, \binits{S.P.}},
\bauthor{\bsnm{Maritan}, \binits{A.}}:
\batitle{Spatial scaling in model plant communities}.
\bjtitle{Phys. Rev. Lett.}
\bvolume{95},
\bfpage{098101}
(\byear{2005})
\doiurl{10.1103/PhysRevLett.95.098101}
\end{barticle}
\endbibitem

%%% 41
\bibitem[\protect\citeauthoryear{Antal et~al.}{2006}]{Antal2006}
\begin{barticle}
\bauthor{\bsnm{Antal}, \binits{T.}},
\bauthor{\bsnm{Redner}, \binits{S.}},
\bauthor{\bsnm{Sood}, \binits{V.}}:
\batitle{Evolutionary dynamics on degree-heterogeneous graphs}.
\bjtitle{Phys. Rev. Lett.}
\bvolume{96},
\bfpage{188104}
(\byear{2006})
\doiurl{10.1103/PhysRevLett.96.188104}
\end{barticle}
\endbibitem

%%% 42
\bibitem[\protect\citeauthoryear{{Ghaffari} and
  {Stollenwerk}}{2012}]{Ghaffari2012}
\begin{bchapter}
\bauthor{\bsnm{{Ghaffari}}, \binits{P.}},
\bauthor{\bsnm{{Stollenwerk}}, \binits{N.}}:
\bctitle{{Evolution of N-species Kimura/voter models towards criticality, a
  surrogate for general models of accidental pathogens}}.
In: \beditor{\bsnm{{Simos}}, \binits{T.E.}},
\beditor{\bsnm{{Psihoyios}}, \binits{G.}},
\beditor{\bsnm{{Tsitouras}}, \binits{C.}},
\beditor{\bsnm{{Anastassi}}, \binits{Z.}} (eds.)
\bbtitle{Numerical Analysis and Applied Mathematics ICNAAM 2012: International
  Conference of Numerical Analysis and Applied Mathematics}.
\bsertitle{American Institute of Physics Conference Series},
vol. \bseriesno{1479},
pp. \bfpage{1331}--\blpage{1334}
(\byear{2012}).
\doiurl{10.1063/1.4756401}
\end{bchapter}
\endbibitem

%%% 43
\bibitem[\protect\citeauthoryear{Caridi et~al.}{2013}]{CARIDI2013216}
\begin{barticle}
\bauthor{\bsnm{Caridi}, \binits{I.}},
\bauthor{\bsnm{Nemiña}, \binits{F.}},
\bauthor{\bsnm{Pinasco}, \binits{J.P.}},
\bauthor{\bsnm{Schiaffino}, \binits{P.}}:
\batitle{Schelling-voter model: An application to language competition}.
\bjtitle{Chaos, Solitons \& Fractals}
\bvolume{56},
\bfpage{216}--\blpage{221}
(\byear{2013})
\doiurl{10.1016/j.chaos.2013.08.013}
\end{barticle}
\endbibitem

%%% 44
\bibitem[\protect\citeauthoryear{Gastner et~al.}{2018}]{Gastner_2018}
\begin{barticle}
\bauthor{\bsnm{Gastner}, \binits{M.T.}},
\bauthor{\bsnm{Oborny}, \binits{B.}},
\bauthor{\bsnm{Gulyás}, \binits{M.}}:
\batitle{Consensus time in a voter model with concealed and publicly expressed
  opinions}.
\bjtitle{Journal of Statistical Mechanics: Theory and Experiment}
\bvolume{2018}(\bissue{6}),
\bfpage{063401}
(\byear{2018})
\doiurl{10.1088/1742-5468/aac14a}
\end{barticle}
\endbibitem

%%% 45
\bibitem[\protect\citeauthoryear{Redner}{2019}]{Redner19}
\begin{barticle}
\bauthor{\bsnm{Redner}, \binits{S.}}:
\batitle{Reality-inspired voter models: A mini-review}.
\bjtitle{Comptes Rendus Physique}
\bvolume{20}(\bissue{4}),
\bfpage{275}--\blpage{292}
(\byear{2019})
\doiurl{10.1016/j.crhy.2019.05.004}
\end{barticle}
\endbibitem

%%% 46
\bibitem[\protect\citeauthoryear{{Garofalo, Jacopo A.} et~al.}{2025}]{GLR25}
\begin{barticle}
\bauthor{\bsnm{{Garofalo, Jacopo A.}}},
\bauthor{\bsnm{{Lippiello, Eugenio}}},
\bauthor{\bsnm{{Rippa, Fabrizio}}}:
\batitle{Competition between long-range and short-range interactions in the
  voter model for opinion dynamics}.
\bjtitle{Eur. Phys. J. B}
\bvolume{98}(\bissue{3}),
\bfpage{49}
(\byear{2025})
\doiurl{10.1140/epjb/s10051-025-00900-x}
\end{barticle}
\endbibitem

%%% 47
\bibitem[\protect\citeauthoryear{Corberi et~al.}{2024}]{Corberi_2024}
\begin{barticle}
\bauthor{\bsnm{Corberi}, \binits{F.}},
\bauthor{\bsnm{Russo}, \binits{S.}},
\bauthor{\bsnm{Smaldone}, \binits{L.}}:
\batitle{Ordering kinetics with long-range interactions: interpolating between
  voter and ising models}.
\bjtitle{Journal of Statistical Mechanics: Theory and Experiment}
\bvolume{2024}(\bissue{9}),
\bfpage{093206}
(\byear{2024})
\doiurl{10.1088/1742-5468/ad6976}
\end{barticle}
\endbibitem

%%% 48
\bibitem[\protect\citeauthoryear{Dyson}{1969}]{Dyson1969}
\begin{barticle}
\bauthor{\bsnm{Dyson}, \binits{F.J.}}:
\batitle{Existence of a phase-transition in a one-dimensional ising
  ferromagnet}.
\bjtitle{Communications in Mathematical Physics}
\bvolume{12}(\bissue{2}),
\bfpage{91}--\blpage{107}
(\byear{1969})
\doiurl{10.1007/BF01645907}
\end{barticle}
\endbibitem

%%% 49
\bibitem[\protect\citeauthoryear{Mukamel}{2009}]{Mukamel2009}
\begin{bchapter}
\bauthor{\bsnm{Mukamel}, \binits{D.}}:
\bctitle{Statistical mechanics of systems with long range interactions}.
In: \beditor{\bsnm{Dauxois}, \binits{T.}},
\beditor{\bsnm{Ruffo}, \binits{S.}},
\beditor{\bsnm{Cugliandolo}, \binits{L.F.}},
\beditor{\bsnm{Arimondo}, \binits{E.}} (eds.)
\bbtitle{Long-Range Interacting Systems}.
\bsertitle{Lecture Notes in Physics},
vol. \bseriesno{970},
pp. \bfpage{3}--\blpage{47}.
\bpublisher{Springer}, \blocation{???}
(\byear{2009}).
\doiurl{10.1063/1.2839123}
\end{bchapter}
\endbibitem

%%% 50
\bibitem[\protect\citeauthoryear{Bar and Mukamel}{2014}]{Bar2014}
\begin{barticle}
\bauthor{\bsnm{Bar}, \binits{A.}},
\bauthor{\bsnm{Mukamel}, \binits{D.}}:
\batitle{Mixed-order phase transition in a one-dimensional model}.
\bjtitle{Physical review letters}
\bvolume{112}(\bissue{1}),
\bfpage{015701}
(\byear{2014})
\end{barticle}
\endbibitem

%%% 51
\bibitem[\protect\citeauthoryear{Aizenman et~al.}{1988}]{Aizenman1988}
\begin{barticle}
\bauthor{\bsnm{Aizenman}, \binits{M.}},
\bauthor{\bsnm{Chayes}, \binits{J.T.}},
\bauthor{\bsnm{Chayes}, \binits{L.}},
\bauthor{\bsnm{Newman}, \binits{C.M.}}:
\batitle{Discontinuity of the magnetization in one-dimensional 1/¦ x- y¦ 2
  ising and potts models}.
\bjtitle{Journal of Statistical Physics}
\bvolume{50}(\bissue{1}),
\bfpage{1}--\blpage{40}
(\byear{1988})
\end{barticle}
\endbibitem

%%% 52
\bibitem[\protect\citeauthoryear{Cardy}{1981}]{Cardy1981}
\begin{barticle}
\bauthor{\bsnm{Cardy}, \binits{J.L.}}:
\batitle{One-dimensional models with 1/r2 interactions}.
\bjtitle{Journal of Physics A: Mathematical and General}
\bvolume{14}(\bissue{6}),
\bfpage{1407}
(\byear{1981})
\end{barticle}
\endbibitem

%%% 53
\bibitem[\protect\citeauthoryear{Redner}{2001}]{bookRedner}
\begin{bbook}
\bauthor{\bsnm{Redner}, \binits{S.}}:
\bbtitle{A Guide to First-Passage Processes}.
\bpublisher{Cambridge University Press},
\blocation{Cambridge}
(\byear{2001})
\end{bbook}
\endbibitem

%%% 54
\bibitem[\protect\citeauthoryear{Cannas et~al.}{2001}]{cannas}
\begin{barticle}
\bauthor{\bsnm{Cannas}, \binits{S.A.}},
\bauthor{\bsnm{Stariolo}, \binits{D.A.}},
\bauthor{\bsnm{Tamarit}, \binits{F.A.}}:
\batitle{Dynamics of ferromagnetic spherical spin models with power law
  interactions: exact solution}.
\bjtitle{Physica A: Statistical Mechanics and its Applications}
\bvolume{294}(\bissue{3-4}),
\bfpage{362}--\blpage{374}
(\byear{2001})
\end{barticle}
\endbibitem

%%% 55
\bibitem[\protect\citeauthoryear{Jedrzejewski
  et~al.}{2015}]{PhysRevE.92.052105}
\begin{barticle}
\bauthor{\bsnm{Jedrzejewski}, \binits{A.}},
\bauthor{\bsnm{Chmiel}, \binits{A.}},
\bauthor{\bsnm{Sznajd-Weron}, \binits{K.}}:
\batitle{Oscillating hysteresis in the $q$-neighbor ising model}.
\bjtitle{Phys. Rev. E}
\bvolume{92},
\bfpage{052105}
(\byear{2015})
\doiurl{10.1103/PhysRevE.92.052105}
\end{barticle}
\endbibitem

%%% 56
\bibitem[\protect\citeauthoryear{Chmiel et~al.}{2018}]{Chmiel2018}
\begin{barticle}
\bauthor{\bsnm{Chmiel}, \binits{A.}},
\bauthor{\bsnm{Gradowski}, \binits{T.}},
\bauthor{\bsnm{Krawiecki}, \binits{A.}}:
\batitle{q-neighbor ising model on random networks}.
\bjtitle{International Journal of Modern Physics C}
\bvolume{29},
\bfpage{1850041}
(\byear{2018})
\doiurl{10.1142/S0129183118500419}
\end{barticle}
\endbibitem

\end{thebibliography}

\end{document}